\documentclass[usenatbib,useAMS,a4paper,fleqn]{mnras}
\usepackage{amssymb}
\usepackage{newtxtext,newtxmath}
\usepackage[T1]{fontenc}
\usepackage{ae,aecompl}
\usepackage{times}
\usepackage{microtype}
\usepackage{verbatim}

\usepackage[normalem]{ulem}

\usepackage{graphicx}
\date{Accepted XXX. Received YYY; in original form ZZZ}
\pubyear{2018}

\usepackage[multidot]{grffile}

\begin{document}

\label{firstpage}
\pagerange{\pageref{firstpage}--\pageref{lastpage}}

\title[Effect of collisions on transiting systems]
  {The dynamical evolution of transiting planetary systems 
    including a realistic collision prescription}

\author[Mustill, Davies \& Johansen]
       {Alexander J. Mustill$^1$\thanks{E-mail: alex@astro.lu.se},
         Melvyn B. Davies$^1$,
         Anders Johansen$^1$\\
         $^1$Lund Observatory, Department of Astronomy \& Theoretical Physics,
         Lund University, Box 43, SE-221 00 Lund, Sweden
       }

\maketitle

\begin{abstract}
Planet--planet collisions are a common outcome of instability in systems 
of transiting planets close to the star, as well as occurring during \emph{in-situ} 
formation of such planets from embryos. 
Previous $N$-body studies of instability amongst 
transiting planets have assumed that collisions result in perfect merging. Here, we 
explore the effects of implementing a more realistic collision prescription on 
the outcomes of instability and \emph{in-situ} formation at orbital radii of a few 
tenths of an au. There is a strong effect on the outcome of the growth of 
planetary embryos, so long as the debris thrown off in collisions is rapidly removed 
from the system (which happens by collisional 
processing to dust, and then removal by radiation 
forces) 
and embryos are small ($<0.1\mathrm{\,M}_\oplus$). 
If this is the case, then systems form fewer detectable 
($\ge1\mathrm{\,M}_\oplus$) planets than 
systems evolved under the assumption of perfect merging in collisions. This 
provides some contribution to the ``\emph{Kepler} Dichotomy'': the observed 
over-abundance of single-planet systems. The 
effects of changing the collision prescription on unstable mature systems of 
super-Earths are less pronounced. 
Perfect mergers only account for a minority of collision outcomes in such 
systems, but most collisions resulting in mass loss are grazing impacts in which 
only a few per cent.\ of mass is lost. As a result, there is little impact on the 
final masses and multiplicities of the systems after instability when compared 
to systems evolved under the assumption that collisions always result in perfect 
merging.
\end{abstract}


\section{Introduction}

%\textcolor{white}{Lorem ipsum this is whitespace to fix a really annoying hyperref error}

\begin{figure}
  \includegraphics[width=0.5\textwidth]{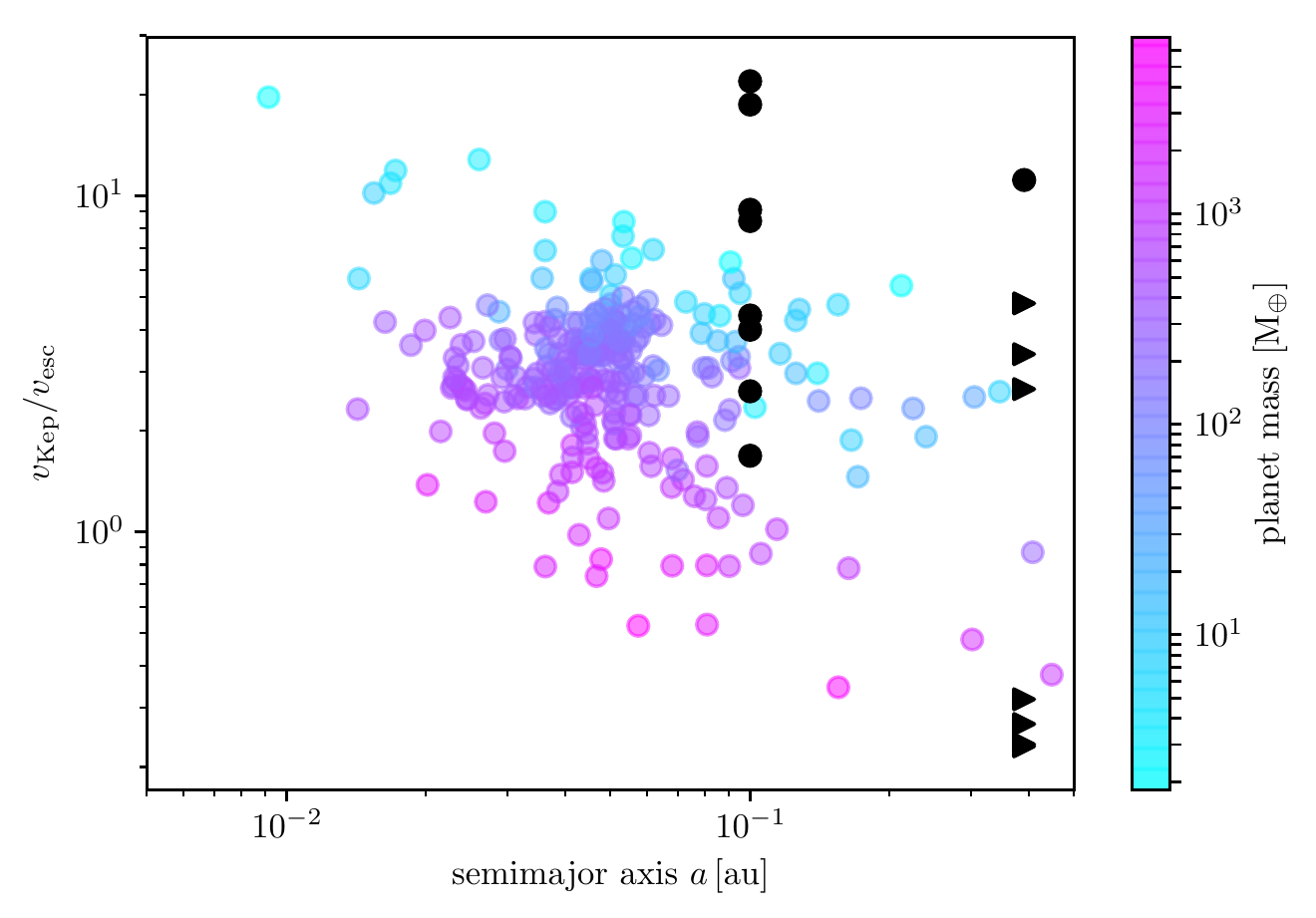}
  \caption{Ratio of Keplerian orbital velocity $v_\mathrm{Kep}$
    to surface escape velocity $v_\mathrm{esc}$ for
    the 272 planets with known masses and radii listed in
    \url{http://exoplanets.org/}. The higher this ratio, the greater
    the chance that a planet--planet collision will be erosive.
    Symbols are coloured
    according to planet mass. The values for Solar System planets are
    marked as black symbols at the right of the plot 
    (Mercury at its semimajor axis as a circle, 
    the other planets as triangles).
    The black symbols at $0.1$\,au show the equivalent
    values for Solar System planets if they were orbiting at
    $0.1$\,au. Super-Earths orbiting at $0.1$\,au are in a
    more erosive regime than the Earth and Venus in our Solar
    System.
  }
  \label{fig:vkep}
\end{figure}

An important challenge in modelling the formation and 
long-term dynamical evolution of planetary systems is 
to adequately model planet--planet collisions. 
Particularly at the short orbital periods 
probed by transit observatories such as 
\textit{Kepler,} when planets collide their impact 
velocity can significantly exceed their surface escape 
velocity. The ratio of orbital to escape velocity for 
known exoplanets with measured masses and radii 
is shown in Figure~\ref{fig:vkep}\footnote{Data 
from \url{http://exoplanets.org/} \citep{Han+14} on 2017-04-25.}. 
We also mark this ratio for the Solar System planets, and 
what the ratio would be if those planets were located at $0.1$\,au. 
The ratio of Keplerian to escape velocity increases towards 
smaller planetary masses. Super-Earths or terrestrial planets 
within $\sim0.1$\,au have $v_\mathrm{Kep}/v_\mathrm{esc}\gtrsim10$, 
comparable to the value for our Solar System's Mercury. 

The resulting high impact velocities between planets 
or planetary embryos means that collisions do not result 
in prefect merging of the parent planets. 
A high-velocity impact may be responsible for the 
anomalously large core of the planet Mercury itself, 
mantle material having been blown off in the impact 
\citep[e.g.,][]{Benz+88,Benz+07}. Amongst the 
population of extrasolar planets discovered by 
\emph{Kepler}, it has been suggested
that the effects of high-velocity collisions may be
responsible for planets in the same system showing
large density differences \citep[e.g., Kepler-36;][]{Quillen+13},
and for contributing to the multiplicity distribution of
close-in transiting planetary systems \citep[e.g.,][]{VolkGladman15}. 
Debris produced in such collisions may be responsible 
for unusually bright warm debris discs 
\citep[e.g.,][]{Song+05,Melis+10,TheissenWest17}.

However, many $N$-body integrators such as \textsc{Mercury} 
\citep{Chambers99} implement collisions between 
bodies as perfect inelastic mergers, with two planets 
combining into one with no loss of mass or momentum 
(but indeed a loss of energy). In constrast, high-velocity collisions 
between planets and planetary embryos in reality can result in 
significant amounts of mass loss. In this paper we explore the effects of adopting a 
collision prescription more realistic than perfect merging 
on the observable multiplicities of 
planetary systems. We compare the final multiplicities 
of unstable multi-planet systems evolved under the assumptions 
of perfect merging with those evolved with a more realistic 
collision prescription. We conduct three experiments. First 
we set up unstable systems of planets at small orbital 
radii ($a<1$\,au), and explore the effects of changing the 
collision prescription on the final multiplicities and 
masses of the planets. We then consider systems of transiting 
planets destabilised by outer planets experiencing 
planet--planet scattering or Kozai perturbations, 
as in our previous paper \citep{Mustill+17}. We show 
that the latter set has a much greater sensitivity to the 
collision prescription, largely owing to the greater eccentricities 
excited in these systems. Finally, we consider the effects of 
changing the collision algorithm on \emph{in-situ} formation 
of rocky super-Earths.

The outcomes of high-velocity collisions between planets 
can be studied numerically, for example with Lagrangian smooth-particle 
hydrodynamics (SPH) or Eulerian adaptive mesh refinement (AMR) codes. 
Different algorithms or resolutions however often result in 
quantitatively different outcomes, even in so fundamental a 
property as the mass of the largest remnant after the collision 
\citep{Genda+15,LiuAgnor+15,ReinhardtStadel17}. Nevertheless, the 
qualitative nature of collision outcomes is understood, with low 
collision velocities (relative to the mutual escape velocity) leading 
to perfect merging, higher velocities leading to some mass loss 
(which may result in growth or erosion of the largest body), and very 
high velocities resulting in ``supercatastrophic disruption'' where 
$<10\%$ of the original mass remains bound. Analytical scaling laws 
determining the boundaries between the different regimes and the 
outcomes in terms of remnant masses have been developed 
\citep{BenzAsphaug99,Genda+12,LeinhardtStewart12,Movshovitz+16}. 
These analytical laws provide an acceptable means to 
compute collision outcomes, particularly given the lack of 
agreement or convergence in some hydrodynamical studies 
\citep[e.g.,][]{Genda+15,LiuAgnor+15,ReinhardtStadel17}. They 
also allow the outcome to be computed near-instantaneously, 
rather than requiring expensive hydrodynamical simulations to 
be conducted for each set of collision parameters. They are 
therefore ideally suited for incorporation into the $N$-body 
integrators used to study planet formation and stability, and 
several studies have now moved away from the perfect merging 
collision model to better model planet formation in the 
terrestrial planet region at $\sim1$\,au 
\citep{Chambers13,Carter+15,Leinhardt+15,Chambers16,Quintana+16} 
or interior \citep*{Wallace+17}.

In this paper we adapt the collision model presented by 
\cite{LeinhardtStewart12}, henceforth LS12, incorporate it into the 
\textsc{Mercury} integrator \citep{Chambers99}, and study the 
effects of this on unstable close-in planetary systems 
detectable in transit at a few tenths of an au. 
We briefly introduce the expected collision environment 
for transiting planetary systems in Section~\ref{sec:colls}. 
We describe the collision model in Section~\ref{sec:model}. 
We then describe the impacts of changing the collision 
model on systems purely composed of close-in planets 
in Section~\ref{sec:inner}, and then on systems of 
close-in planets destabilised by outer planets in 
Section~\ref{sec:outer}. We then examine the effects on 
\emph{in situ} formation of rocky super-Earths in 
Section~\ref{sec:formation}. We discuss and conclude in 
Section~\ref{sec:discussion}.

\section{The collision environment of close-in planets}

\label{sec:colls}

\begin{figure*}
  \includegraphics[width=0.53\textwidth]{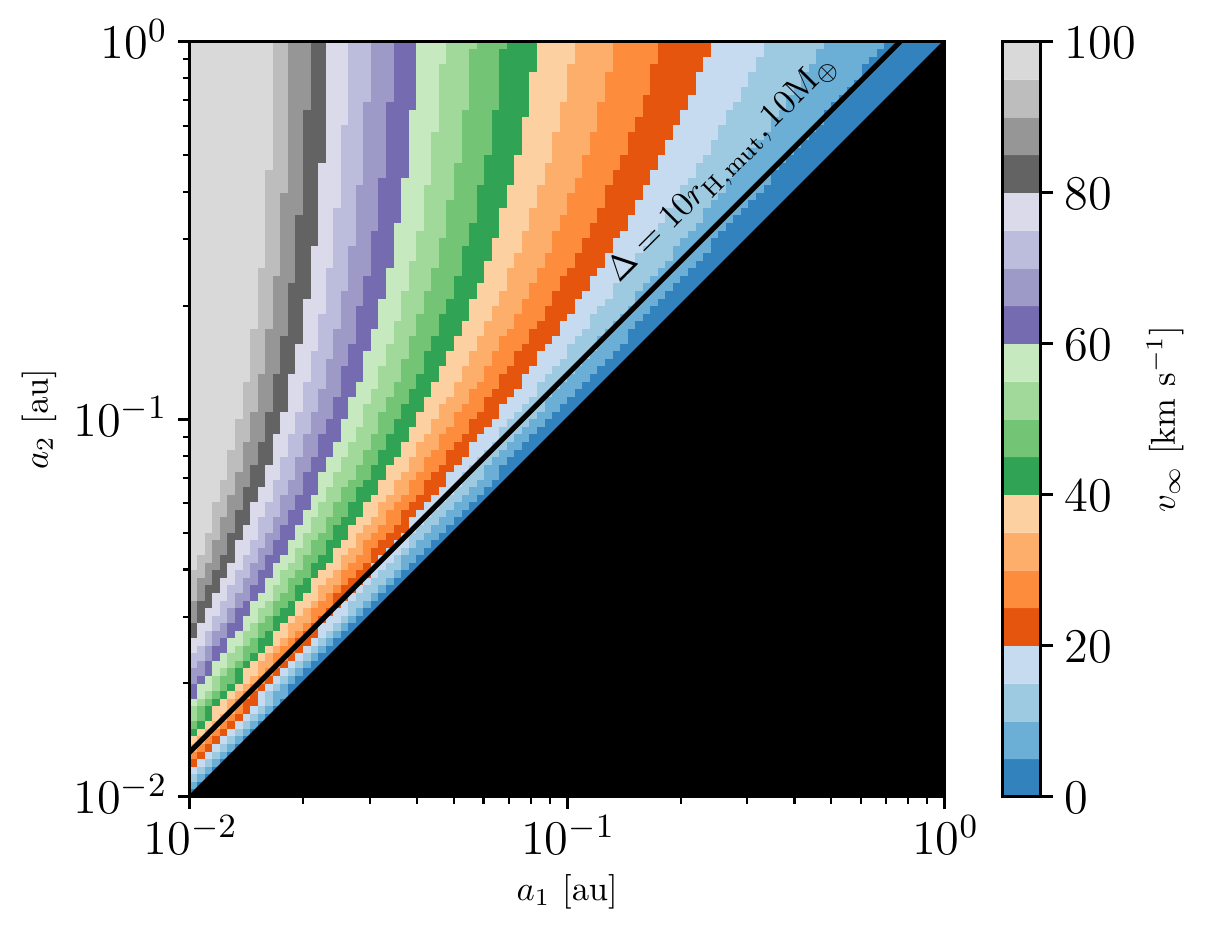}
  \includegraphics[width=0.42\textwidth]{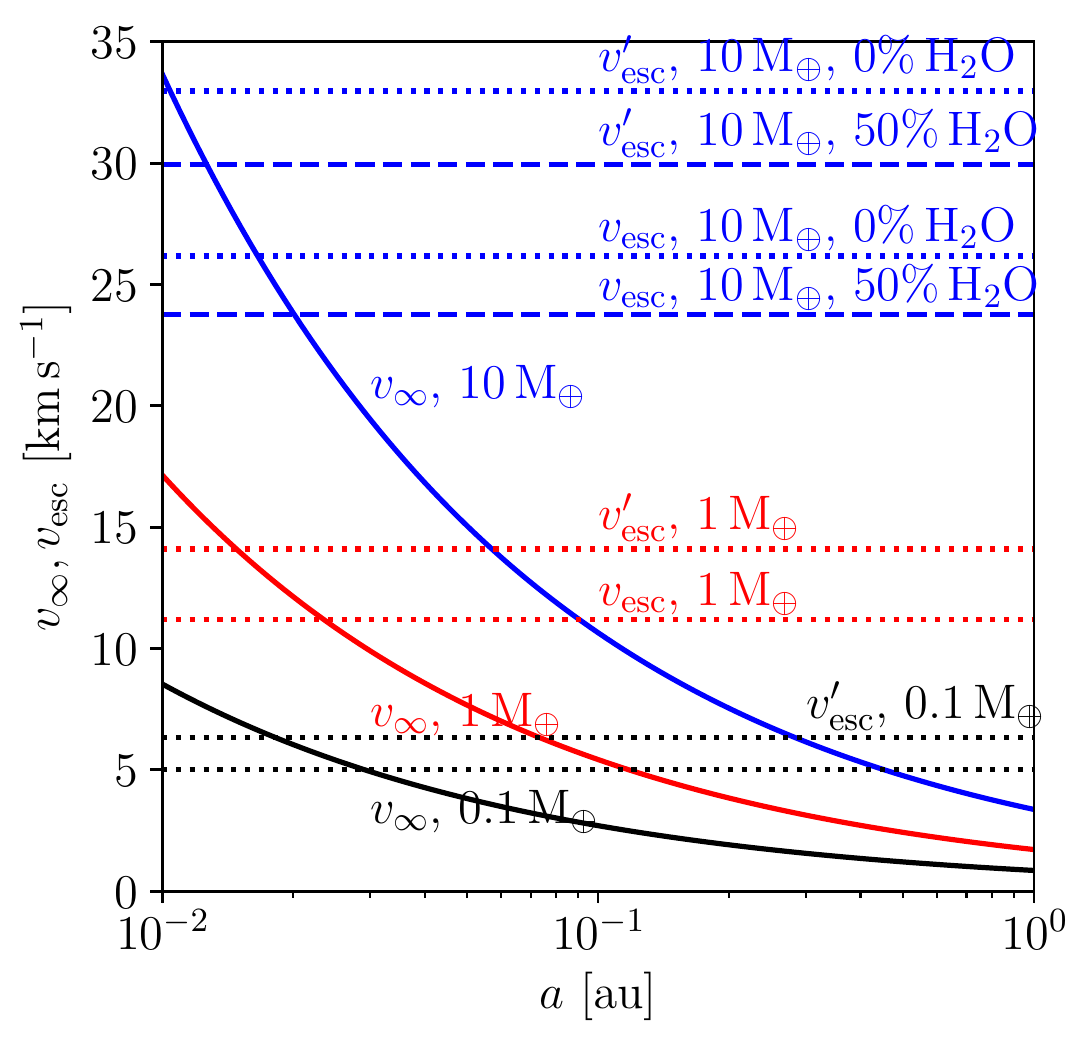}
  \caption{\textbf{Left: }Relative velocity at infinity if an outer planet at semimajor 
    axis $a_2$ collides at pericentre with an inner 
    planet on a coplanar circular orbit at $a_1$. We also show 
    the line $a_2=a_1+10r_\mathrm{H,mut}$ to mark 
    the semimajor axis of an outer planet spaced 10 mutual 
    Hill radii from the inner, when both planets 
    have masses $10\mathrm{\,M}_\oplus$. 
    \textbf{Right: }Relative velocity at infinity for 
    two planets spaced 10 mutual Hill radii when the planet masses 
    are both $0.1$, $1$ or $10\mathrm{\,M}_\oplus$ (solid lines). We also 
    show (dotted/dashed lines) the single-planet surface escape velocity 
    $v_\mathrm{esc}$ and the mutual escape velocity 
    $v_\mathrm{esc}^\prime$, for Mars-like composition for 
    $0.1\mathrm{\,M}_\oplus$ planets, Earth-like composition for $1\mathrm{\,M}_\oplus$
    planets and for two compositions for $10\mathrm{\,M}_\oplus$ planets. 
    Encounter velocities at infinity 
    are a significant fraction of the escape velocity at 
    $a\lesssim1$\,au. Gravitational focusing will boost the actual 
    impact velocity significantly above escape velocity.}
  \label{fig:vrel}
\end{figure*}

We begin by estimating the collision velocities that can be expected 
between transiting planets within a few tenths of an au. These must be 
compared to the planets' surface escape velocities: the more the 
collision velocity exceeds the escape velocity, the more energy is available 
to unbind a fraction of the total mass, and the worse an approximation 
perfect merging is to the true collision outcome.

First consider 
one planet on a circular orbit at semimajor axis $a_1$, with 
a second planet on a wider coplanar orbit of semimajor axis $a_2$ 
and pericentre $q_2=a_1$, so that its orbit just touches that of 
the first planet and collisions can occur at pericentre. The orbital 
Keplerian velocity of the inner planet is 
\begin{equation}
  v_\mathrm{Kep,1} = \sqrt{\mathcal{G}M_\star/a_1},
\end{equation}
while the pericentre velocity of the outer planet is 
\begin{equation}
  v_\mathrm{peri,2} = \sqrt{\mathcal{G}M_\star\left(1+e_2\right)/a_1},
\end{equation}
$M_\star$ being the mass of the star and $e_2$ the eccentricity of the outer 
planet. Treating the encounter between the two planets as a two-body 
scattering event, the relative velocity ``at infinity'' (i.e., close to the 
pericentre of the outer planet, before the planets' mutual 
gravity becomes significant) is therefore 
\begin{equation}
  v_\infty = \sqrt{\mathcal{G}M_\star\left(1+e_2\right)/a_1} - 
  \sqrt{\mathcal{G}M_\star/a_1}.
\end{equation}
This velocity is shown in the space of $\left(a_1,a_2\right)$ 
(where $a_1<a_2$) in the left-hand panel of Figure~\ref{fig:vrel}. 
In extreme cases, of the inner planet being located at a few 
hundredths of an au, extreme relative velocities of 
$v_\infty>100\mathrm{\,km\,s}^{-1}$ are attainable. For planets 
at $0.1$\,au, $v_\infty>20\mathrm{\,km\,s}^{-1}$ can be 
attained. This is already almost twice Earth's escape velocity 
($v_{\mathrm{esc,}\oplus}\approx11\mathrm{\,km\,s}^{-1}$). 
Note that the velocity estimated here is a \emph{minimum} for 
the $v_\infty$ of the planets' encounter: the addition of 
radial or vertical velocity components (if the pericentre 
of the outer planet's orbit lies inside the orbit of the 
inner planet, or if there is a mutual orbital inclination) 
will increase $v_\infty$ further.

How widely-separated may the orbits of planets initially be 
while still permitting eventual collision? This depends greatly 
on the underlying cause of instability in the system. In this Paper 
we consider two classes of system: tightly-packed systems 
which are intrinsically unstable, and intrinsically stable 
inner systems which are destabilised by the excitation of 
eccentricities in outer bodies (beyond $\sim1$\,au) by 
planet--planet scattering or Kozai perturbations from a 
binary companion. 
In the former case, consider 
several planets separated by $\Delta r_\mathrm{H,mut}$, where $\Delta$ 
is the spacing measured in units of the mutual Hill radius
\begin{equation}
  r_\mathrm{H,mut} = \frac{a_1+a_2}{2}\left(\frac{M_1+M_2}{3M_\star}\right)^{1/3},
\end{equation}
where $M_i$ are the masses of the two planets. Instability in these systems 
is thought to be driven by the overlap of three-body mean motion resonances 
\citep{Quillen11}, and the timescale for instability is a sensitive 
function of $\Delta$ 
\citep[e.g.,][]{Chambers+96,FaberQuillen07,SmithLissauer09,Mustill+14}, 
and systems of five Earth-mass planets are unstable within a few billion 
orbits at least out to separations of $\Delta=9$ \citep{SmithLissauer09}. 
We mark the $a_2$ corresponding to a separation $\Delta=10$ for 
$10\mathrm{\,M}_\oplus$ planets as a solid black line on the left-hand 
panel of Figure~\ref{fig:vrel}. We find \emph{minimum} encounter 
velocities $v_\infty$ of over $30\mathrm{\,km\,s}^{-1}$ at a 
few hundredths of an au. In the right-hand panel of Figure~\ref{fig:vrel} 
we show this encounter velocity as a function of $a_1$, for Mars-mass, 
Earth-mass, and $10\mathrm{\,M}_\oplus$ planets. We also 
show the escape velocity of these planets in two forms: the single-planet 
surface escape velocity 
\begin{equation}
  v_\mathrm{esc} = \sqrt{2\mathcal{G}M/R},
\end{equation}
which is appropriate for the escape of a massless particle from 
the planet's surface, and the 2-planet mutual escape velocity 
as defined by LS12
\begin{equation}
  v_\mathrm{esc}^\prime = \sqrt{2\mathcal{G}M^\prime/R^\prime}
\end{equation}
where $M^\prime$ is the combined mass of the two planets and 
$R^\prime$ is the radius of a sphere of the same total mass and 
density as the two planets. This latter is the appropriate measure 
for determining the outcome of a collision between comparable-mass 
bodies, and for equal-mass, equal-density planets is greater than the 
single-planet escape velocity by a factor $2^{1/3}$. 
For the calculation of the escape velocity of the $10\mathrm{\,M}_\oplus$ 
planets, we use the mass--radius relation of \cite{Valencia+07} for 
two compositions ($0\%$ water and $50\%$ water); the Mars-mass 
and Earth-mass planets have escape velocities of their 
Solar System archetypes. The minimum 
$v_\infty$ exceeds the escape velocity within $\sim0.02$\,au. 
However, as mentioned previously, the impact velocity will be 
enhanced by additional components of the velocity vector and 
by gravitational focusing. Numerical experiments by 
\cite{VolkGladman15} found that in intrinsically unstable 
systems such as those we are considering, the collision 
velocities (particularly those following the first collision) 
easily exceed twice the escape velocity.

Systems destabilised by external influences, however, can 
be expected to be excited to higher eccentricities and therefore 
higher encounter velocities. The terrestrial planets of the 
Solar System have a small chance of experiencing an 
instability driven by the $\nu_5$ secular resonance 
\citep{LaskarGastineau09,Batygin+15}. In this case, Mercury can be 
placed on a collision trajectory with Venus: a separation of 
$0.336$\,au, or 63 mutual Hill radii, requiring an eccentricity 
on Mercury's part of $e=0.87$ if Venus retains a circular orbit. 
We have previously shown \citep{Mustill+15,Mustill+17} that systems 
of transiting planets can be destabilised by the action of 
dynamically-excited planets beyond 1\,au. In these case, 
collisions can occur between widely-separated transiting inner planets, or between 
one of the inner planets and one of the outer planets if its 
eccentricity is sufficiently highly excited, placing these systems 
towards the upper left of the left-hand panel of Figure~\ref{fig:vrel} 
with very high impact velocities of $\sim100\mathrm{\,km\,s}^{-1}$ 
to be expected, particularly if orbits become highly inclined or 
even retrograde. Other sources 
of instability that may result in high-velocity collisions 
would be in systems destabilised by secular inclination resonance sweeping, 
for example as a rapidly-rotating young star spins down 
\citep{SpaldingBatygin16}, where the inclination excitation 
could also lead to high relative velocities.

These arguments show that the collisional environment of 
close-in planets is extreme, and that the collisions that 
occur will often be significantly higher than the planets' 
escape velocity. Therefore, the usual assumption of 
perfect merging in $N$-body integrators will no longer be 
valid. We now proceed to numerically study how common outcomes 
other than perfect merging are, with a modified \textsc{Mercury} 
$N$-body code. We describe the implementation of the collision 
algorithm in the next section, before proceeding to study 
the collision statistics in intrinsically unstable systems, 
and then those destabilised by outer bodies. Finally, we 
study the effects on \emph{in-situ} formation of rocky 
transiting planets.

\section{The collision model}

\label{sec:model}

\begin{figure}
  \includegraphics[width=.5\textwidth]{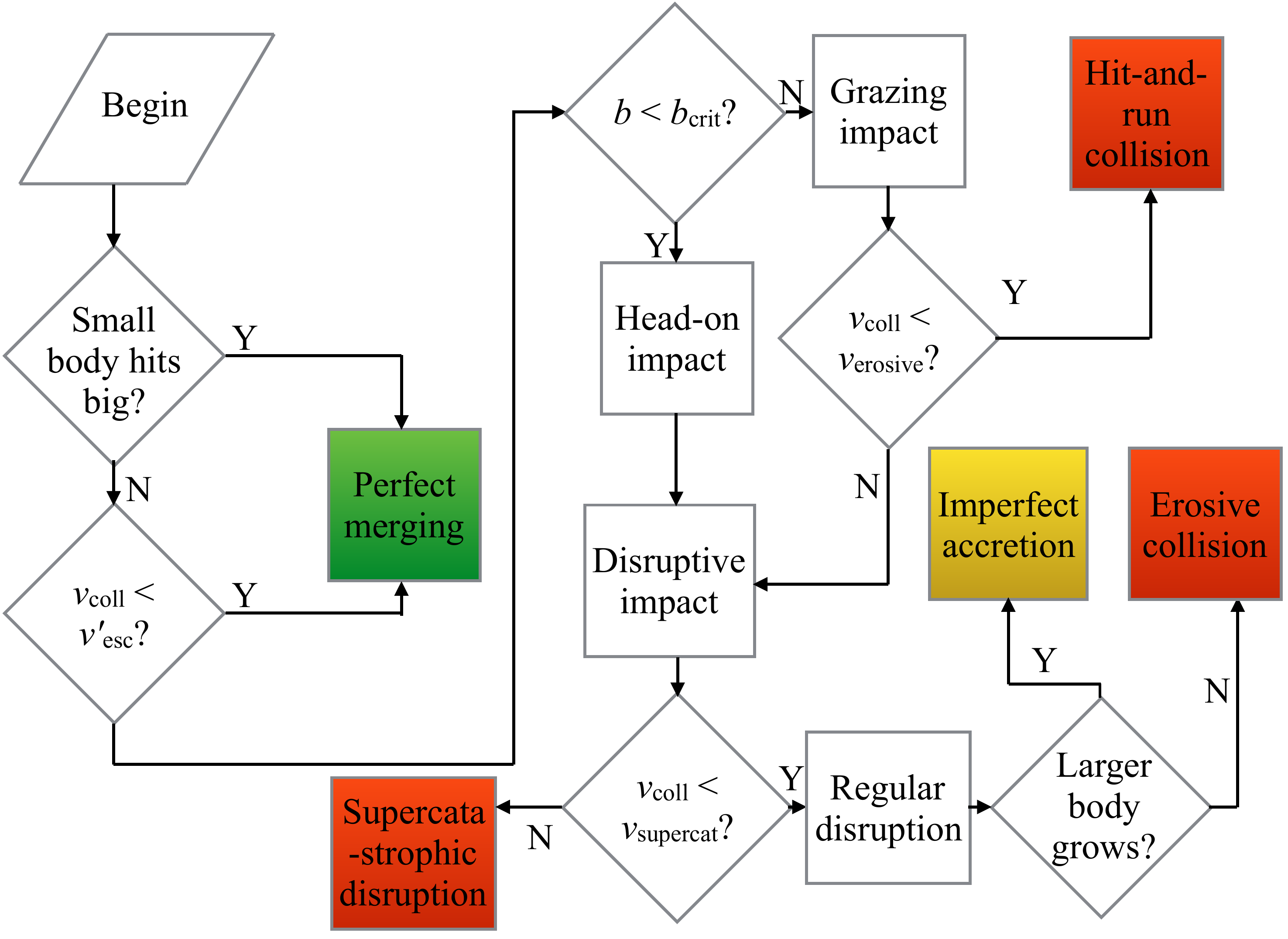}
  \caption{Flowchart illustrating the collision algorithm 
    we implemented in the \textsc{Mercury} code, 
    based on \protect\cite{LeinhardtStewart12}. If the collision 
    velocity is lower than the mutual escape velocity, 
    the standard \textsc{Mercury} perfect merging routines are 
    called, whereby the bodies merge into one with the combined 
    mass of both the original bodies; 
    this also occurs if a small body in \textsc{Mercury} 
    (a massive test particle) strikes a big body. 
    If the collision velocities exceed this threshold, we check for 
    grazing collisions (if the impact parameter is above a critical 
    value, so that the centre of mass of the smaller body misses the larger). 
    These lead to a hit-and-run outcome with minimal 
    mass loss, if the impact velocity is below the erosion threshold. 
    Head-on impacts or higher-velocity grazing 
    impacts lead to disruptive outcomes: here the larger body may 
    still grow, with some mass lost to collisional debris 
    (imperfect accretion), it may lose mass (erosion), or it may 
    experience a supercatastrophic collision at very high 
    impact velocities. See Section~\ref{sec:algorithm} for further details. 
    The colour coding of outcomes (green for perfect merging, yellow for imperfect 
    accretion, and red for erosion) corresponds to the bars in 
    Figures~\ref{fig:dm}, \ref{fig:outcomes-i0.1} and \ref{fig:dm-outer}.}
  \label{fig:algorithm}
\end{figure}

We adjust the \textsc{Mercury} $N$-body integrator to better 
handle high-velocity collisions in the following way. First, 
we alter the routines checking for a collision between 
bodies. In \textsc{Mercury} this check is performed 
by an interpolation of particle trajectories within 
each time-step to see whether an intersection has taken 
place. We need to know, in addition, what the parameters 
of the collision are (collision velocity and impact parameter), not 
merely whether or not one has occurred. Therefore, if a collision 
is detected, we rerun the integration from the start of that timestep 
for half of the duration of the timestep. This is repeated iteratively 
until the timestep falls to $10^{-5}$ days, thus providing accurate 
values for the impact parameter and velocity. As collisions are relatively 
infrequent, the computational cost of this is negligible.

Next we change the outcomes of the collisions themselves. 
We base our algorithm on that of LS12, suitable for collisions 
between rocky protoplanets or planets. We are therefore studying 
genuine ``super-Earths'' rather than ``mini-Neptunes'' which 
have a significant gaseous component, a distinction which seems 
to occur at around $1.5\mathrm{\,R}_\oplus$ 
\citep{WeissMarcy14,Fulton+17}. 
The collision algorithm takes as input 
the masses and radii (and hence densities) of the two bodies, 
and the impact parameter $b$ and the collision velocity $v_\mathrm{coll}$ 
(both of these \emph{at the moment of impact,} not at a point at 
infinity). From these inputs, collisions are classified into 
perfect merging (small $v_\mathrm{coll}$), hit-and-run (large 
$b$ but not enough $v_\mathrm{coll}$ for disruption), disruptive 
(moderately large $v_\mathrm{coll}$; some material lost from 
the two planets, while the larger may accrete some material or be eroded) 
and super-catastrophic (very large $v_\mathrm{coll}$; 
less than $10\%$ of the total mass remains bound in largest remnant). 
The algorithm is described in detail in the next subsection.

In the case of disruptive and hit-and-run collisions, mass is lost 
from one or both of the planets. Some rock is vapourised, and a 
fraction of this will recondense into small dust grains. Other 
ejecta will be thrown off as large, gravitationally-bound chunks. 
We treat the debris as small bodies in \textsc{Mercury}: super-particles 
which have mass, interact gravitationally and collide with the big 
bodies but have no direct interactions with each other. This approach means 
that, as far as purely gravitational forces go, we can neglect 
uncertainties in the size distribution of the debris. However, 
the debris is also subject to collisional processing and radiation 
forces \citep{Burns+79,Wyatt+07,JacksonWyatt12} which can significantly reduce its lifetime. 
Due to the complexities of modelling the initial size distribution 
of collisional debris and its subsequent evolution, we consider two 
extreme cases: for most of our simulations, we simply remove the 
debris instantaneously from the simulations. This represents the case where 
the debris is primarily in the form of small grains or vapour and 
is removed on dynamical time-scales by the stellar wind and radiation forces, 
or else is rapidly collisionally ground down to such small 
sizes. We discuss this in more detail in Section~\ref{sec:fragments}. 
For some simulations, we retain 100\% of the debris as small bodies in 
\textsc{Mercury}. This represents the opposite extreme in which most of 
the mass of the debris is in a few large fragments. Reality will 
lie somewhere between these two scenarios.

\subsection{The collision algorithm}

\label{sec:algorithm}

The collision algorithm we implement is shown schematically in 
Figure~\ref{fig:algorithm} and described below:

\begin{enumerate}
\item \textbf{Remove small bodies involved in a collision.} 
  If a large body collides with a small body (as defined by \textsc{Mercury}: 
  small bodies are massive 
  test particles) then perfect merging is assumed.
\item \textbf{Calculate the interacting mass $m_\mathrm{interact}$:} 
  the mass of the fraction 
  of the projectile whose trajectory intersects the target. 
  This fraction is calculated approximately 
  in LS12, Equation~10. However, we found that this formula was not 
  always accurate, particularly when the projectile and the target 
  were of comparable radii, and we evaluate the interacting 
  mass by numerical quadrature in all cases. We also account for the 
  possibility of a less massive but geometrically larger ``projectile'' of low density.
\item \textbf{Calculate the mutual escape velocity} 
  $v_\mathrm{esc}^\prime=\sqrt{2M^\prime/R^\prime}$ (LS12 Equation~53). 
  We slightly change the definition of $M^\prime$ and $R^\prime$ compared 
  to LS12: we set $M^\prime=M_1+M_2$ not $M_1+m_\mathrm{interact}$ as 
  all the mass of the two bodies interacts gravitationally, and 
  we set $R^\prime=\left(3(M_1/\rho_1+M_2/\rho_2)/(4\pi)\right)^{1/3}$ 
  to account for the projectile and the target potentially having different densities. 
  Defining the mutual escape velocity in this way allows us to treat as perfect mergers 
  hit-and-run impacts that later merge \citep{Genda+12}, although there is a 
  slight dependence on impact parameter that we do not capture.
\item \textbf{If the collision velocity 
  $v_\mathrm{coll}\le v_\mathrm{esc}^\prime$ then a perfect merger occurs. }
  Note that the mutual escape velocity is defined for a spherical configuration 
  of matter, whereas when the two planets make contact they are in a dumbbell-shaped 
  higher-energy configuration. Therefore, two bodies at rest at infinity 
  will not have reached their mutual escape velocity at the moment of impact, 
  and so collisions at velocities less than the mutual escape velocity are possible.
\item \textbf{Determine whether a grazing impact occurs.} If the trajectory of the centre of mass of 
  the less massive body does not intersect the target, the impact is counted as grazing.
%\item \textbf{Calculate the critical impact velocity for disruption $V^{\prime\star}$.} Here 
%  we use Equations 17, 22 and 30 of LS12.
\item \textbf{Calculate the threshold for an erosive collision in which the 
  larger planet loses mass.} We set $M_\mathrm{lr}=M_\mathrm{target}$ in LS12 Equation 5.
\item \textbf{Determine whether a hit-and-run collision has occurred.} This occurs if 
  the impact velocity is below the erosion threshold and the collision is grazing. If 
  a hit-and-run collision occurs then jump to (xii).
\item \textbf{Calculate the threshold for super-catastrophic disruption.} 
  This is given by LS12 Equation 5 with $M_\mathrm{lr}=M_\mathrm{tot}/10$.
\item \textbf{Determine the class of collision outcome: imperfect accretion, erosion 
  or supercatastrophic disruption.} This depends on the impact velocity 
  compared to the computed thresholds.
\item \textbf{Calculate the largest remnant mass.} This is given 
  by LS12 Equation 4 for regular disruption, and LS12 Equation 44 for supercatastrophic 
  disruption. We do not preserve the \textit{second-largest remnant} as 
  a separate big body, since its mass is usually $<10\%$ of the initial mass of 
  the smaller body (LS12 Equation 37).
\item \textbf{Assign new velocity to the largest remnant.} In this we follow 
  the procedure in Section~3.3 of LS12 if the mass of the largest body decreases. 
  If the mass increases, this may not conserve momentum, and we therefore 
  add to the larger body's momentum that of the material accreted from the smaller body.
\item \textbf{Determine outcome of hit-and-run collisions.} The algorithm of 
  LS12 does not specify the amount of mass lost in a hit-and-run collision. However, 
  a small fraction of mass is indeed lost. We approximate this with a geometrical 
  approach where the material in the overlap region calculated in step (ii) is removed. 
  In our fiducial runs decribed below, the median mass removed in a hit-and-run 
  collision was $2\%$, although some result in greater mass loss ($90\%$ lost 
  less than $16\%$).
\item \textbf{Distribute mass lost into small-body fragments if desired.} A fraction 
  $f_\mathrm{remove}$ of the mass lost in the collision is instantaneously removed 
  from the simulation. This is to represent material which is vapourised and, after 
  recondensing, will be removed from the system on a short time-scale by radiation 
  forces. The remainder of the material is distributed into fragments, modelled 
  as small bodies in \textsc{Mercury}. As these are super-particles, we distribute the 
  mass equally amongst them. To prevent an excessive number of fragments 
  being generated and slowing down the simulations, we spawn 10 particles 
  per collision. We set $f_\mathrm{remove}=1$ for most of this paper.
\item \textbf{Assign velocities to fragments. }Following \cite{JacksonWyatt12}, 
  who fit one simulation outcome from \cite{Marcus+09} for the Moon-forming collision, 
  we assign fragments velocities with a truncated Gaussian distribution. 
  We spawn the fragments at the edge of its Hill sphere with an isotropic distribution, 
  and scale the velocity distribution by the planetary escape velocity.
\item \textbf{Adjust velocity of largest remnant to conserve momentum.} If fragments are 
  generated, we adjust the momentum of the largest remnant after assigning velocities 
  to the collision fragments.
\end{enumerate}

\section{Numerical study I: instability amongst the inner planets}

\label{sec:inner}

\subsection{Setup}

\begin{figure*}
  \includegraphics[width=0.32\textwidth]{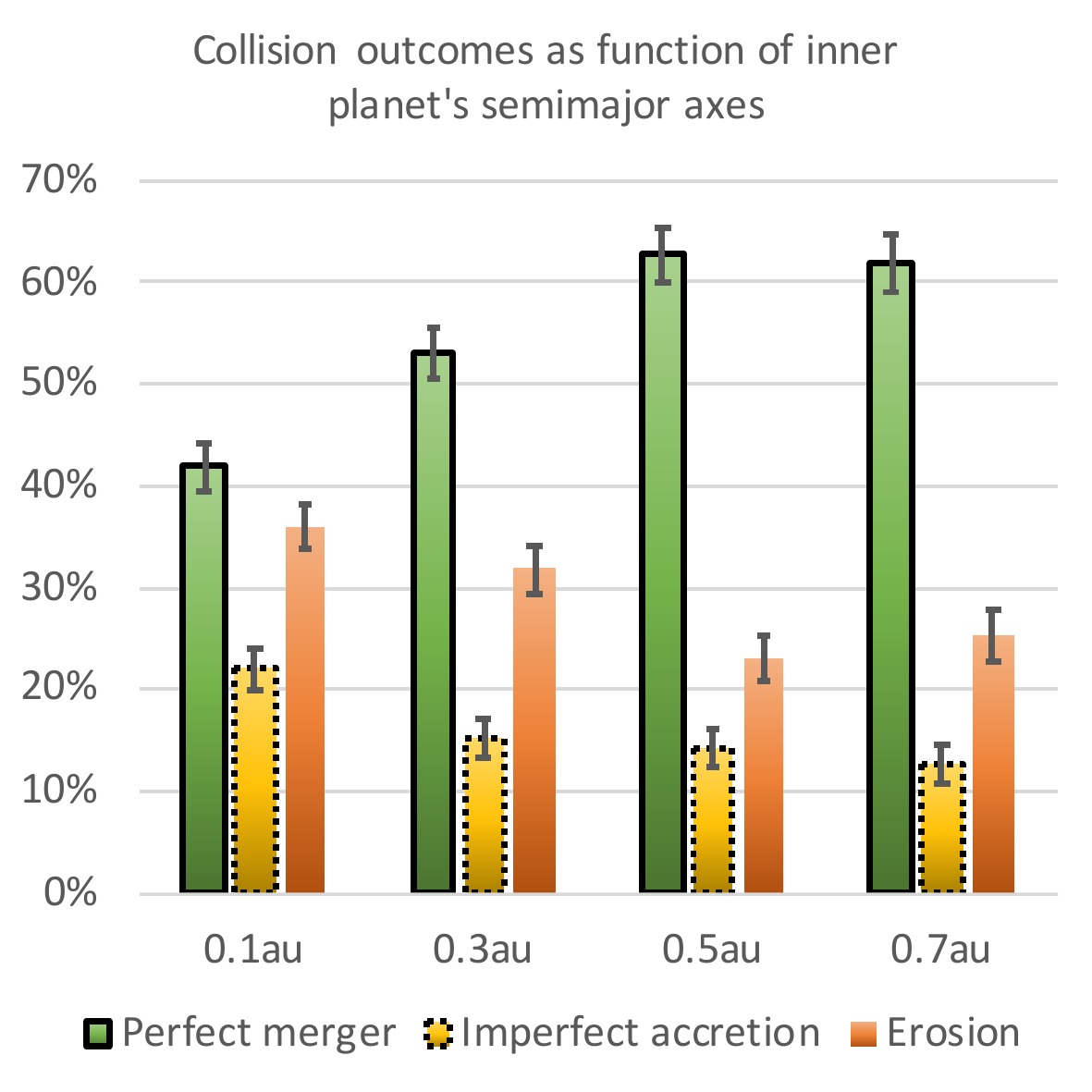}
  \includegraphics[width=0.32\textwidth]{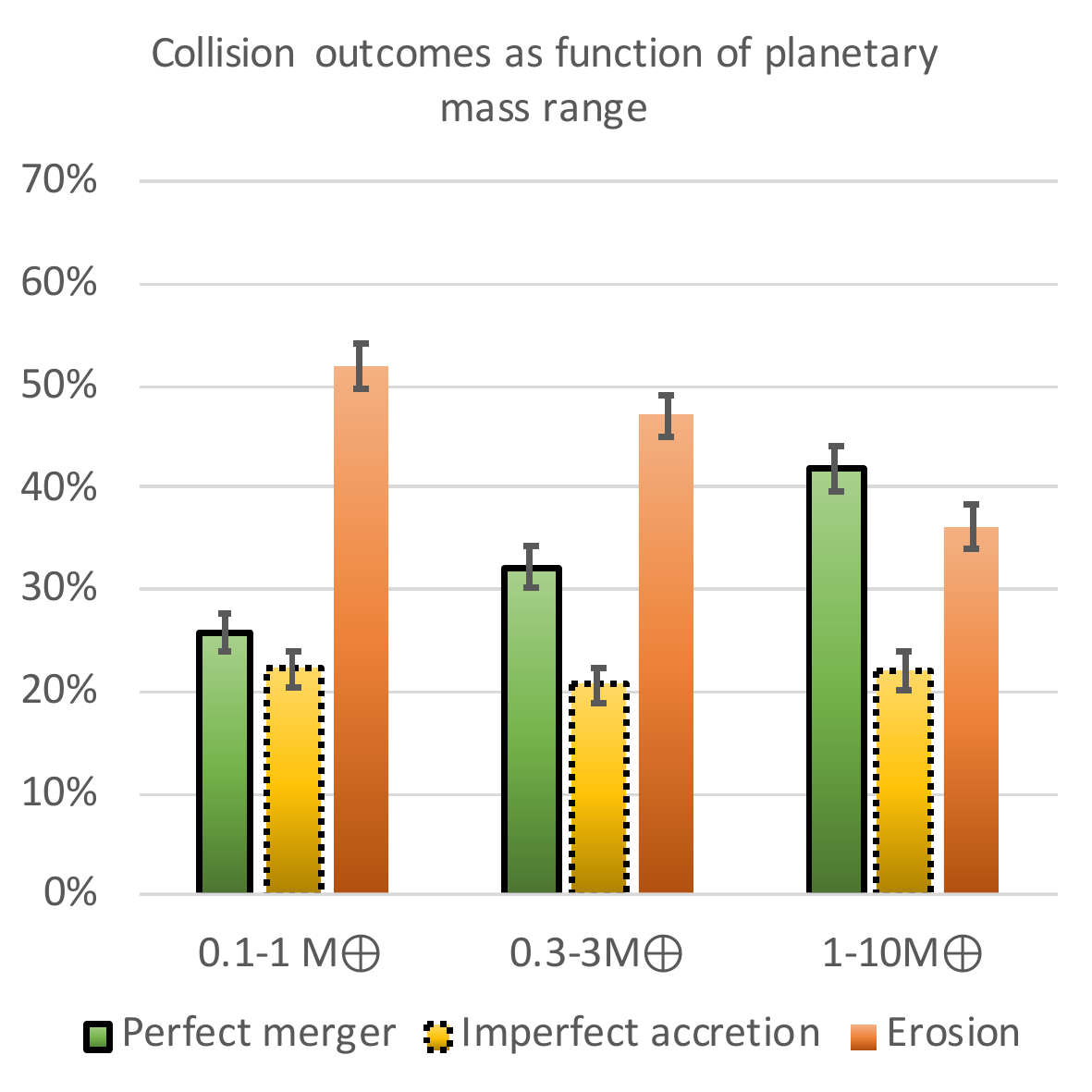}
  \includegraphics[width=0.32\textwidth]{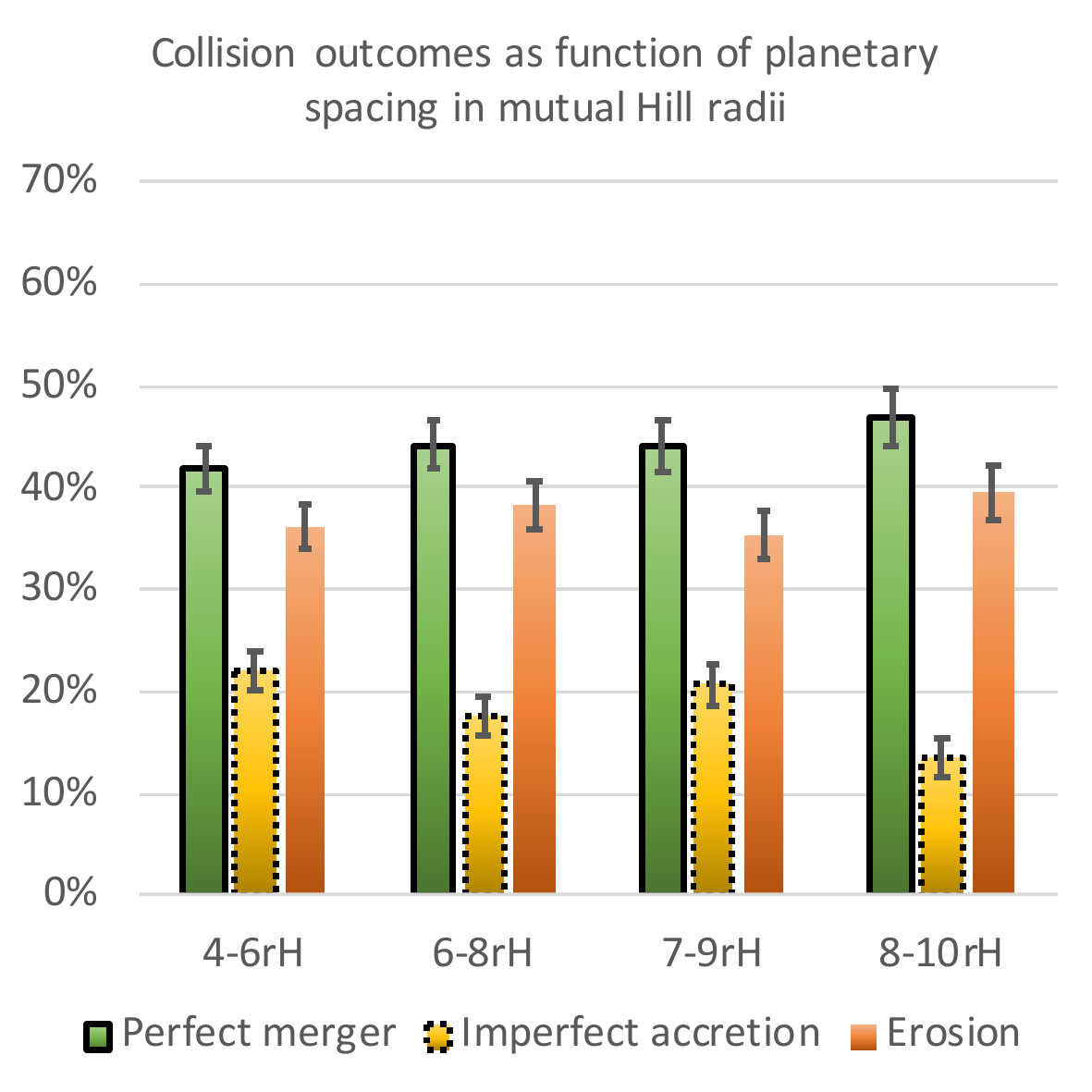}
  \includegraphics[width=0.32\textwidth]{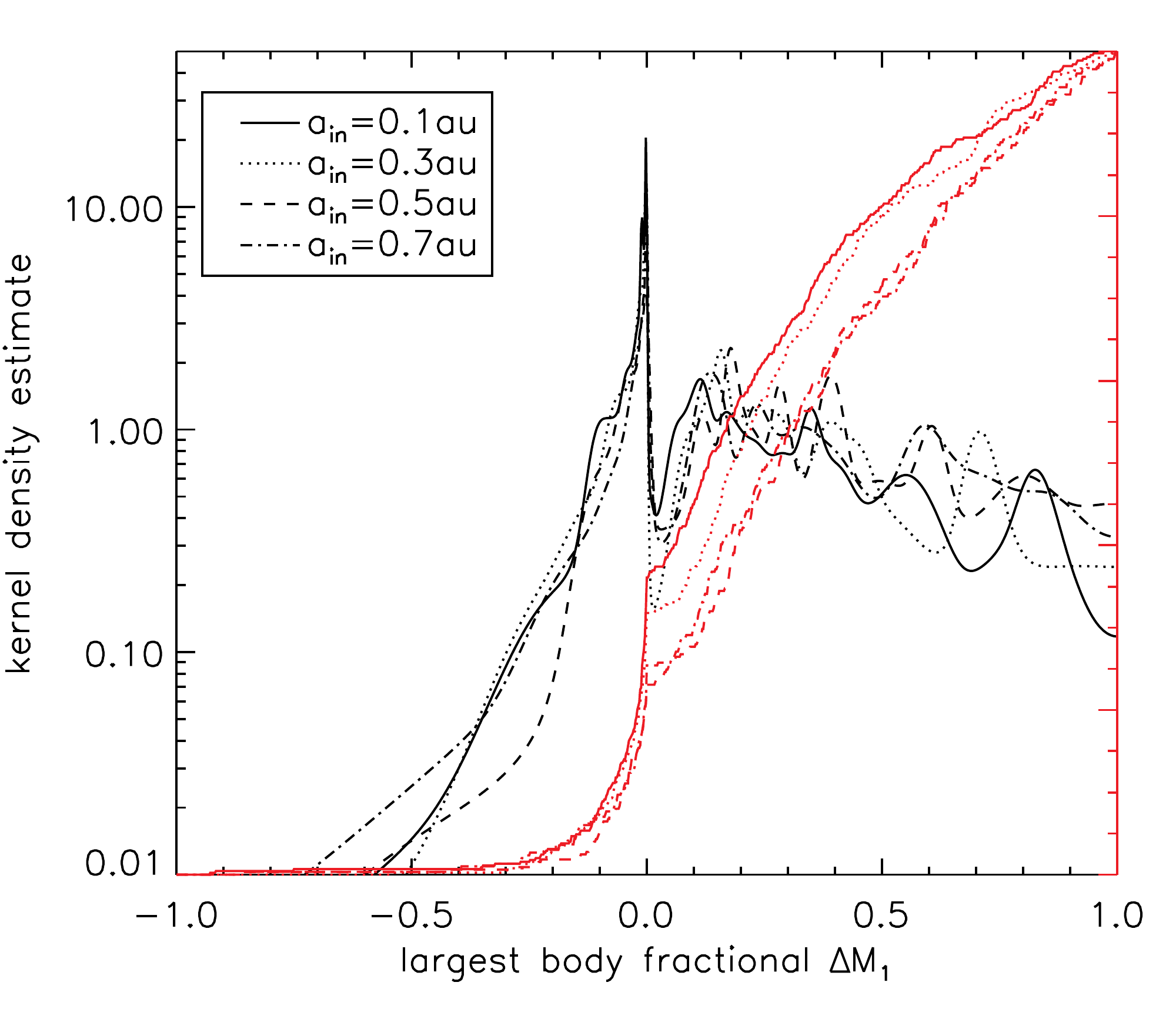}
  \includegraphics[width=0.32\textwidth]{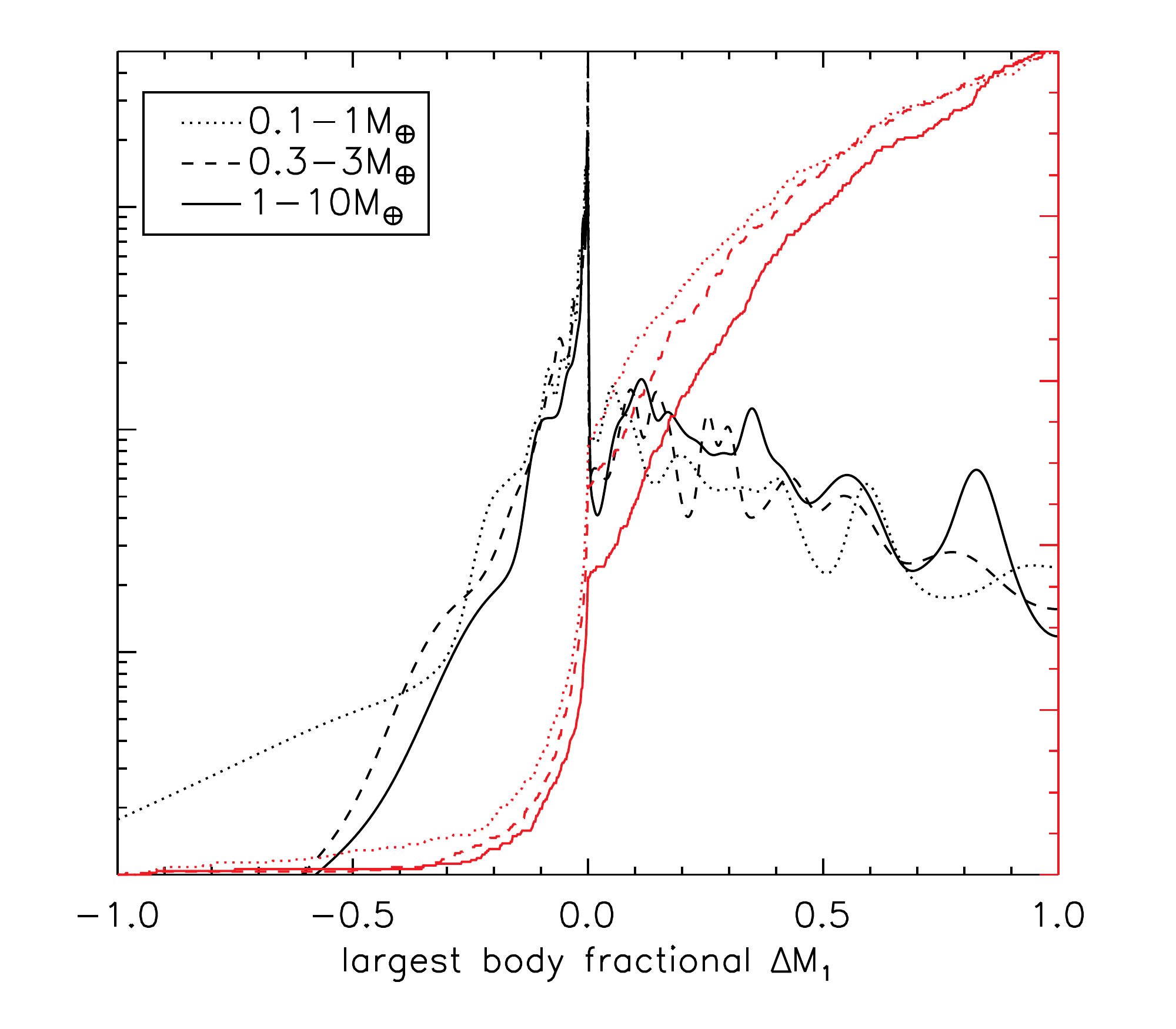}
  \includegraphics[width=0.32\textwidth]{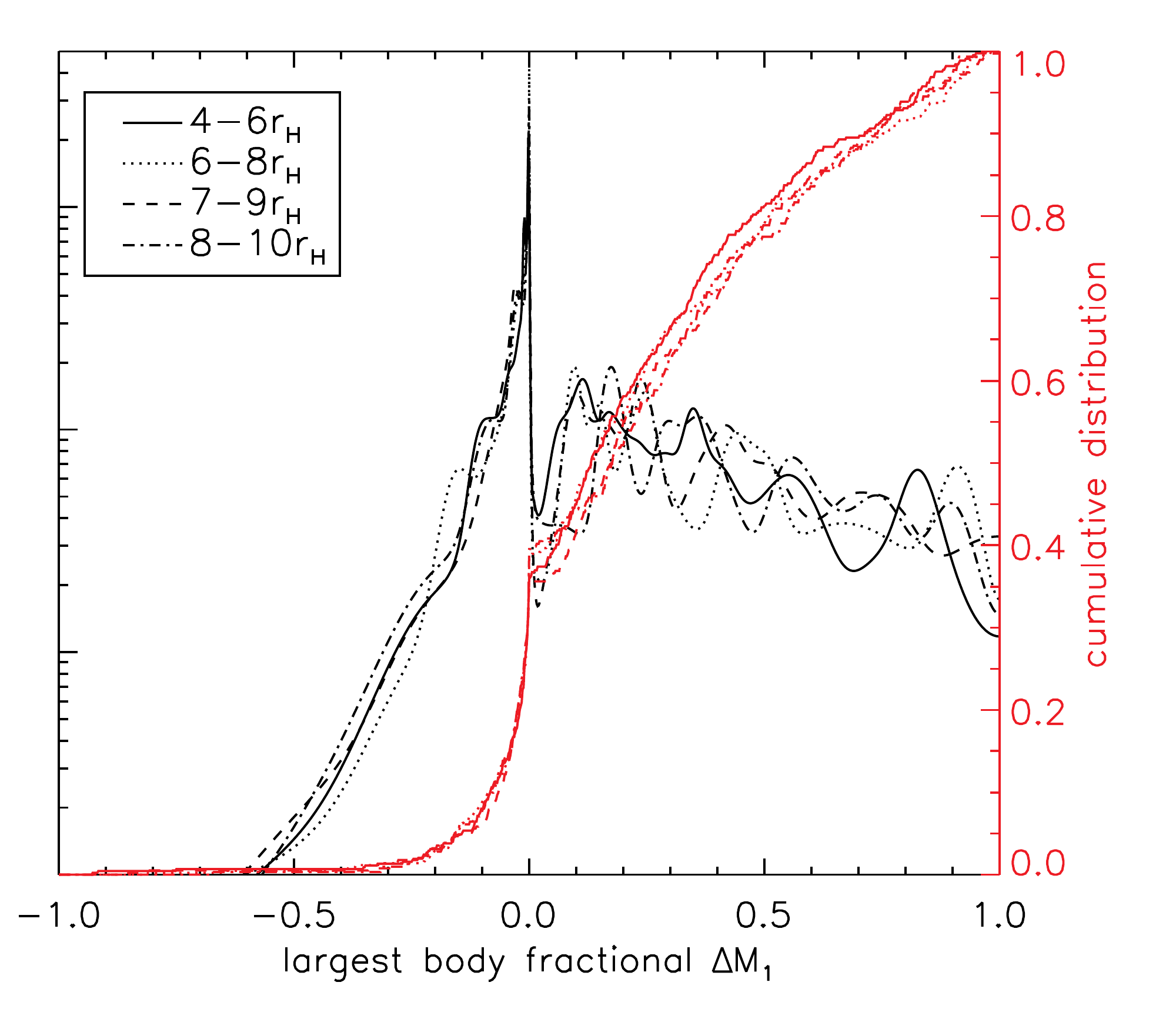}
  \caption{Outcomes of collisions in unstable systems of five planets. 
    The fiducial system has planets with masses $1-10\mathrm{\,M}_\oplus$, 
    spaced by $4-6$ mutual Hill radii, with the innermost planet 
    at $0.1$\,au; we vary one parameter per column of panels. 
    The upper panels show the broad breakdown of collision outcomes, with each 
    set of three bars representing a set of 100 simulations. Collisions are divided into 
    perfect merging (both planets merge with no mass loss), 
    imperfect accretion (larger planet grows but 
    does not accrete 100\% of the mass of the smaller planet), and 
    erosion (larger planet loses some mass; this includes 
    hit-and-run impacts).
    The lower panels then show the distributions of the 
    fractional change in mass for the largest planet 
    in the collision ($\Delta M_1/M_1$). 
    Kernel density estimates are shown in
    black against the left-hand vertical axis, while the cumulative distributions
    are shown in red against the right-hand axis. 
    The large spike of slightly-erosive 
    collisions is largely due to hit-and-run impacts, which result in a small 
    amount of mass loss. 
    \textbf{Left: }The effects of changing the 
    planets' semimajor axes. The innermost planet
    is placed at $0.1$, $0.3$, $0.5$ or $0.7$\,au; masses are $1-10\mathrm{M}_\oplus$.
    We find a higher fraction of erosive collisions closer to the star,
    as expected from the higher Keplerian (hence impact) velocities. However,
    even at $0.1$\,au, most collisions remain accretional.
    \textbf{Centre: }The effects of changing the mass of the planets. The innermost planet 
    is always at $0.1$\,au; masses are in the range
    indicated in the legend. The smaller planets experience a higher fraction
    of erosive collisions, as expected from their lower surface escape velocities.
    \textbf{Right: }The effects of changing the planets' initial separations 
    in mutual Hill radii. 
    The initial separation in mutual Hill radii for the
    planets is set to the ranges $4-6$, $6-8$, $7-9$ and $8-10r_\mathrm{H,mut}$.
    The initial spacing has no noticeable effect on the distribution of
    collision outcomes.
    In each panel there are 100 runs in each set of
    integrations.}
  \label{fig:dm}
\end{figure*}

We first turn to purely internal instabilities, of the kind studied by
\cite{Johansen+12}, \cite{PuWu15}, \cite{VolkGladman15} 
and \citep{Izidoro+17}. We set up
tightly-packed systems of five super-Earths 
(masses drawn logarithmically from $1-10\mathrm{\,M}_\oplus$)
and evolve them for 10\,Myr with both the perfect merging assumption
and the LS12 algorithm. We run four sets with each collision prescription,
with the innermost planet being placed at $0.1$, $0.3$, $0.5$ or $0.7$\,au.
We also run sets with the innermost planet at $0.1$\,au for planetary
mass ranges $0.1-1\mathrm{\,M}_\oplus$ and $0.3-3\mathrm{\,M}_\oplus$. Planets
are separated by $4-6$ mutual Hill radii, which ensures dynamical instability
on short timescales (from a few 100 yrs onwards). We also test more
widely-spaced but still unstable systems to verify that the results are not
sensitive to the initial separation, running simulation sets at $0.1$\,au
with initial separations of $6-8$, $7-9$ and $8-10$ mutual Hill radii. Most 
simulations start with planets' inclinations drawn from $0^\circ$ 
to $5^\circ$ of the reference plane, which provides a typical mutual 
inclination of $\sim3^\circ$ \citep{Johansen+12}. 
To test the effects of the initial inclination distribution, we also run 
simulations from a flatter initial configuration with inclinations drawn 
from $0^\circ$ to $0.1^\circ$.

Our fiducial case has the innermost
planet at $0.1$\,au, masses in the range $1-10\mathrm{\,M}_\oplus$, 
separations from $4-6$ mutual Hill radii, and inclinations up to $5^\circ$.

\subsection{Results}

%\subsubsection{Collision outcomes}

The outcomes of collisions are summarized in Figure~\ref{fig:dm}. The upper panels 
show a broad breakdown of collision outcomes into perfect mergers, 
accretion at less than 100\% efficiency, and erosive or disruptive collisions 
where the mass of the largest planet decreases. The majority of 
erosive collisions are grazing impacts which result in a relatively 
small fraction of mass lost. The lower panels show the 
detailed distributions (as kernel density estimates and cumulative 
distributions) of the mass change in each collision in each simulation 
set, normalised to the mass of the larger planet. In the left-hand column 
we vary the innermost planet's semimajor axis, in the centre 
column the range of planetary masses, and in the right-hand 
column the planetary spacing. 

The upper panels show that most collisions very close to the star 
do not result in perfect merging: only $42\%$ of collisions 
in the $0.1$\,au simulations resulted in perfect merging. We see 
a trend towards more gentle collisions as we move away from the star: 
in the simulations at $0.7$\,au, the fraction of perfect mergers 
has risen to above 60\%. The mass of the planets has a very strong 
effect, with only $26\%$ of collisions resulting in perfect mergers 
when the planetary masses are in the range $0.1-1\mathrm{\,M}_\oplus$. 
These trends are expected from a consideration of planetary 
orbital and escape velocities, as discussed in Section~\ref{sec:colls}: 
in a fixed planet mass range (escape velocity), decreasing the 
distance to the star increases the Keplerian and therefore the 
collision ($v_\infty$) velocities, while at a fixed semimajor axis, decreasing 
the mass decreases the escape velocity but keeps the Keplerian 
velocity constant. Nevertheless, the erosive
nature of collisions between the smaller planets on short-period orbits
 may pose a challenge to
models of \emph{in-situ} formation from lower-mass planetary embryos
\citep[e.g.,][]{HansenMurray12,HansenMurray13,ChatterjeeTan14,
Schlichting14,MoriartyFischer15,Ogihara+15,MoriartyBallard16},
and we investigate this more thoroughly in Section~\ref{sec:formation}.
Finally, we see essentially no dependence of collision outcomes 
on the initial spacing of the planets. Increasing the spacing increases the time-scale 
for the onset of collisions, but it does not affect the velocities 
once collisions begin.

In the lower panels of Figure~\ref{fig:dm} we show in detail the 
distribution of mass changes to the largest planet. While many 
collisions are erosive, many of these are in fact hit-and-run 
collisions resulting in little mass loss, seen as the large spike 
in the kernel density estimate just below $\Delta M=0$. Only a few 
percent of collisions result in the larger planet losing more 
than $10\%$ of its pre-impact mass. In total, $7-11\%$ of the 
total initial mass of the planets in the runs at $0.1$\,au 
with planets in the $1-10$ Earth mass range 
was lost, slightly lower in the more widely-spaced 
systems which experienced fewer collisions. This fraction 
is considerably higher when starting from lower-mass planets: 
$19\%$ in the $0.3-3$ Earth mass range and $25\%$ in the 
$0.1-1$ Earth mass range.

\begin{figure*}
  \includegraphics[width=\textwidth]{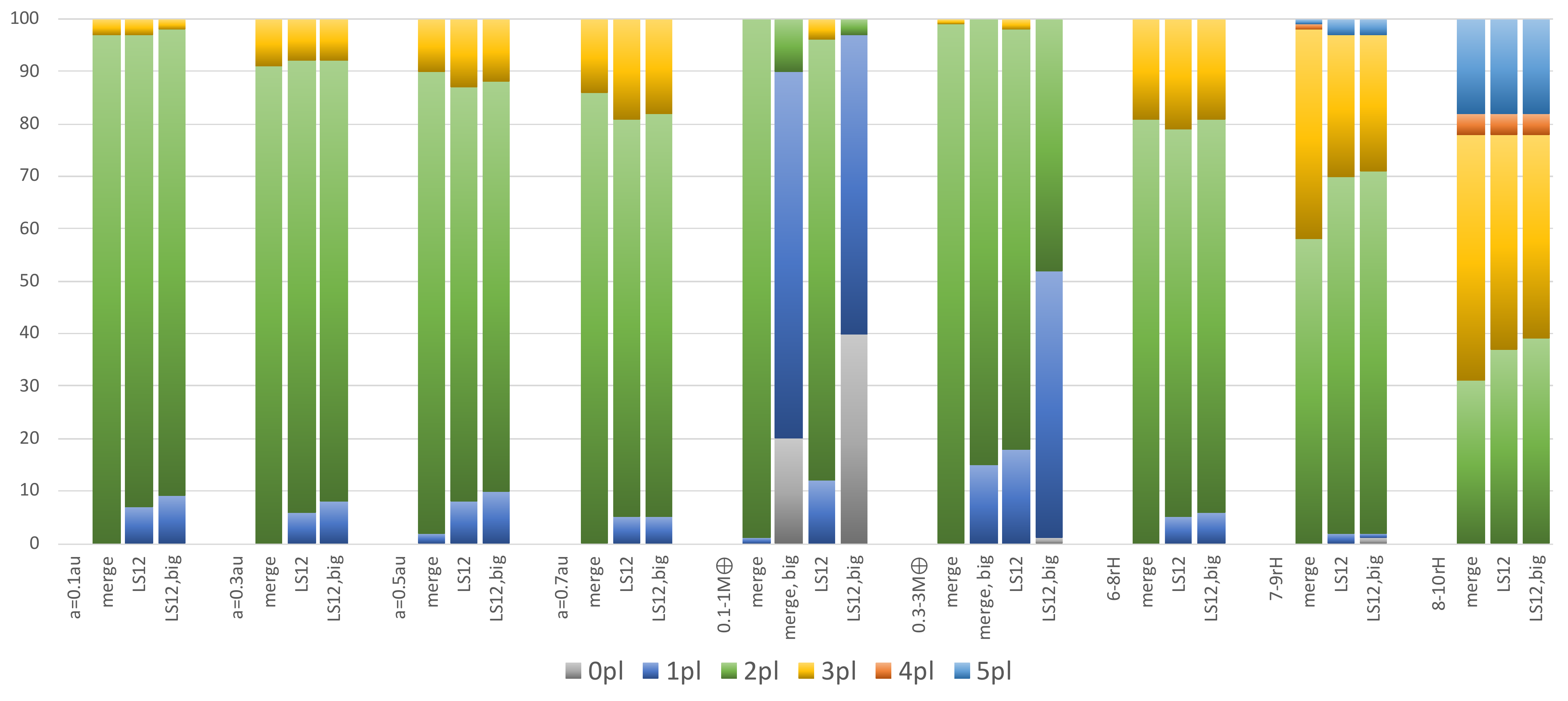}
  \caption{Final multiplicities of the initially 
    5-planet systems of Figure~\ref{fig:dm}. Outcomes are shown 
    as a function of semimajor axis (first four sets), 
    mass (fifth and sixth sets), and initial orbital spacing in 
    mutual Hill radii (final three sets). Each set of simulations 
    is run both with the standard \textsc{Mercury} collision algorithm 
    where collisions always result in perfect mergers (``merge''), 
    and with our more realistic algorithm (``LS12''). In columns 
    labelled only ``merge'' and ``LS12'', we count all planets; in 
    columns labelled ``\ldots, big'' we count only planets 
    of at least $1\mathrm{\,M}_\oplus$. 
    Most systems reduce to 2-planet systems, or 3-planet 
    systems for the initially more widely separated ones. 
    When counting only planets at least as massive as Earth, 
    many zero- or single-planet systems form when the initial 
    planet masses are small. This is exacerbated when using 
    the realistic collision model.}
  \label{fig:npl}
\end{figure*}

In Figure~\ref{fig:npl} we show the final numbers 
of planets in the systems. The systems typically 
reduce to 2-planet systems; the more widely-spaced 
ones have more planets surviving at the end, but 
may reduce further if integrated for longer. For 
the simulation sets where the final planet mass can 
be below $1\mathrm{\,M}_\oplus$ (any with the LS12 
collision model, and any with initial planet masses 
below $1\mathrm{\,M}_\oplus$), we show separately 
the numbers of planets with mass $>1\mathrm{\,M}_\oplus$, 
as a crude criterion for detectability by \emph{Kepler}. 
For most simulation sets, there is little difference 
in the final multiplicities when comparing the 
perfect merging prescription with LS12. This is 
true even when including ``undetectable'' planets 
smaller than $1\mathrm{\,M}_\oplus$. The exceptions to 
this are when starting from smaller planets: we then 
see significant differences in the numbers of planets 
that survive with mass greater than Earths between 
the two collision models: for the $0.1-1\mathrm{\,M}_\oplus$ 
simulations, we find with perfect merging $20\%$ of 
systems having zero detectable planets, $70\%$ having 
one, and $10\%$ having two; with the LS12 algorithm, 
these numbers become respectively $40\%$, $57\%$ and 
$3\%$.

\begin{figure}
  \includegraphics[width=0.45\textwidth]{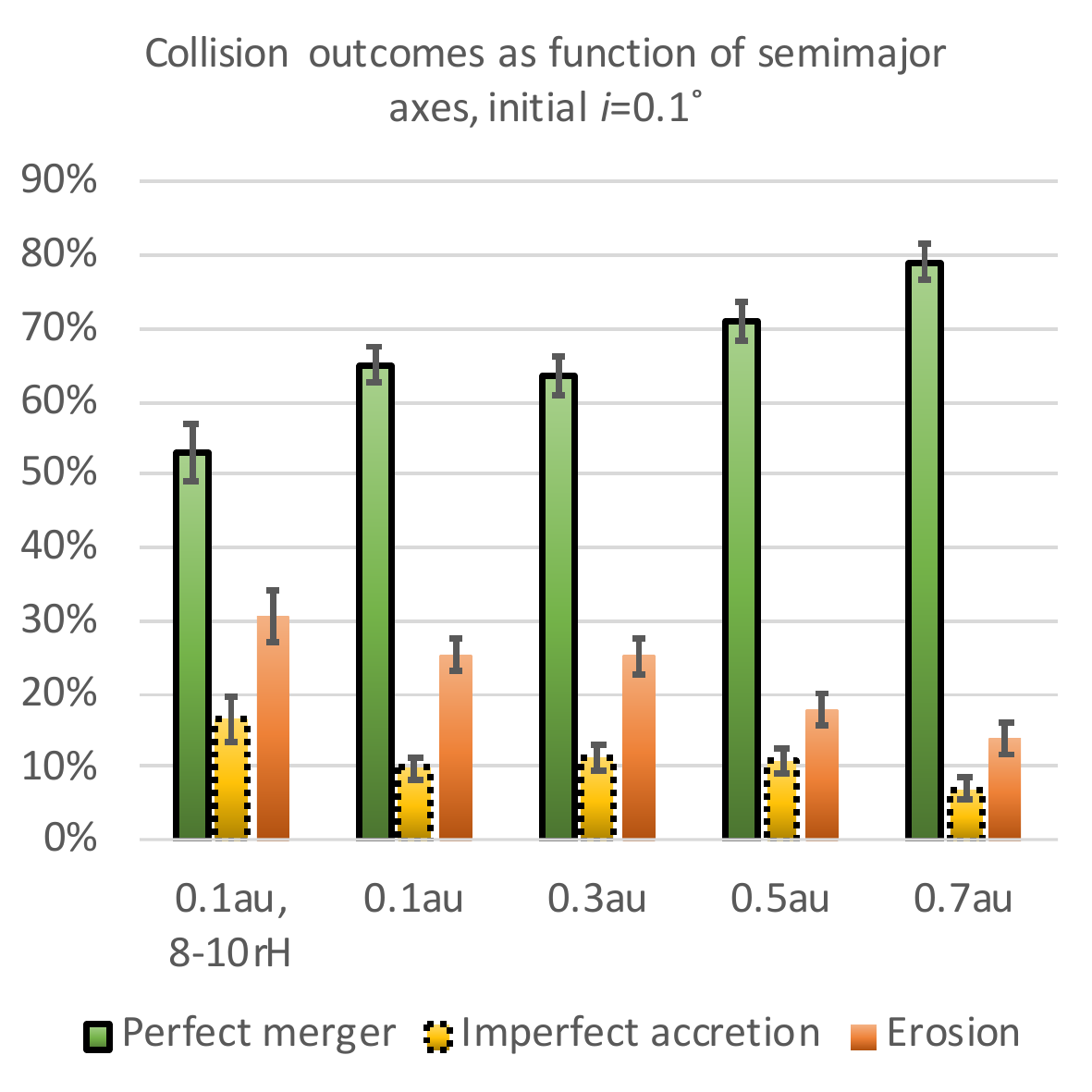}
  \caption{Collision outcomes when starting from a flat 
    configuration with initial inclinations in the range
    $0^\circ$ to $0.1^\circ$ (compare to left-hand panels of
    Figure~\ref{fig:dm}, where the inclinations were up to $5^\circ$). 
    Initial spacings are all $4-6$ mutual
    Hill radii except for the leftmost set, spaced at $8-10$
    mutual Hill radii. In most of these flat systems, 
    a larger fraction of collisions result
    in perfect mergers than in the initially more inclined
    systems shown in Figure~\ref{fig:dm}, although the
    more widely-spaced systems ($8-10r_\mathrm{H,mut}$)
    have statistics comparable to their more inclined counterparts.}
  \label{fig:outcomes-i0.1}
\end{figure}

For the runs discussed so far, we have assumed 
an initial inclination distribution of up to 
$5^\circ$, meaning that the mutual inclinations 
peak at around $3^\circ$ \citep{Johansen+12}. 
We now study the effects of starting from initially 
flatter systems, with inclinations only up to 
$0.1^\circ$ (Figure~\ref{fig:outcomes-i0.1}). 
Here we find a larger fraction of perfect mergers 
as collision outcomes at small orbital radii: 
$65\%$ in the flat systems at $0.1$\,au spaced 
$4-6$ mutual Hill radii, compared to 
$42\%$ in the inclined systems. However, here 
there is a dependence on planetary spacing, and 
for the systems spaced $8-10$ mutual Hill radii 
the fractions more closely resemble the ones 
from the inclined systems: $53\%$ perfect 
mergers and $31\%$ erosive. Here, the wider 
systems have more time in which to become dynamically 
excited, with a final mean inclination 
of $2.6^\circ$ for the most widely-spaced 
set compared to $1.9^\circ$ for the tightest set.

\begin{figure}
  \includegraphics[width=0.5\textwidth]{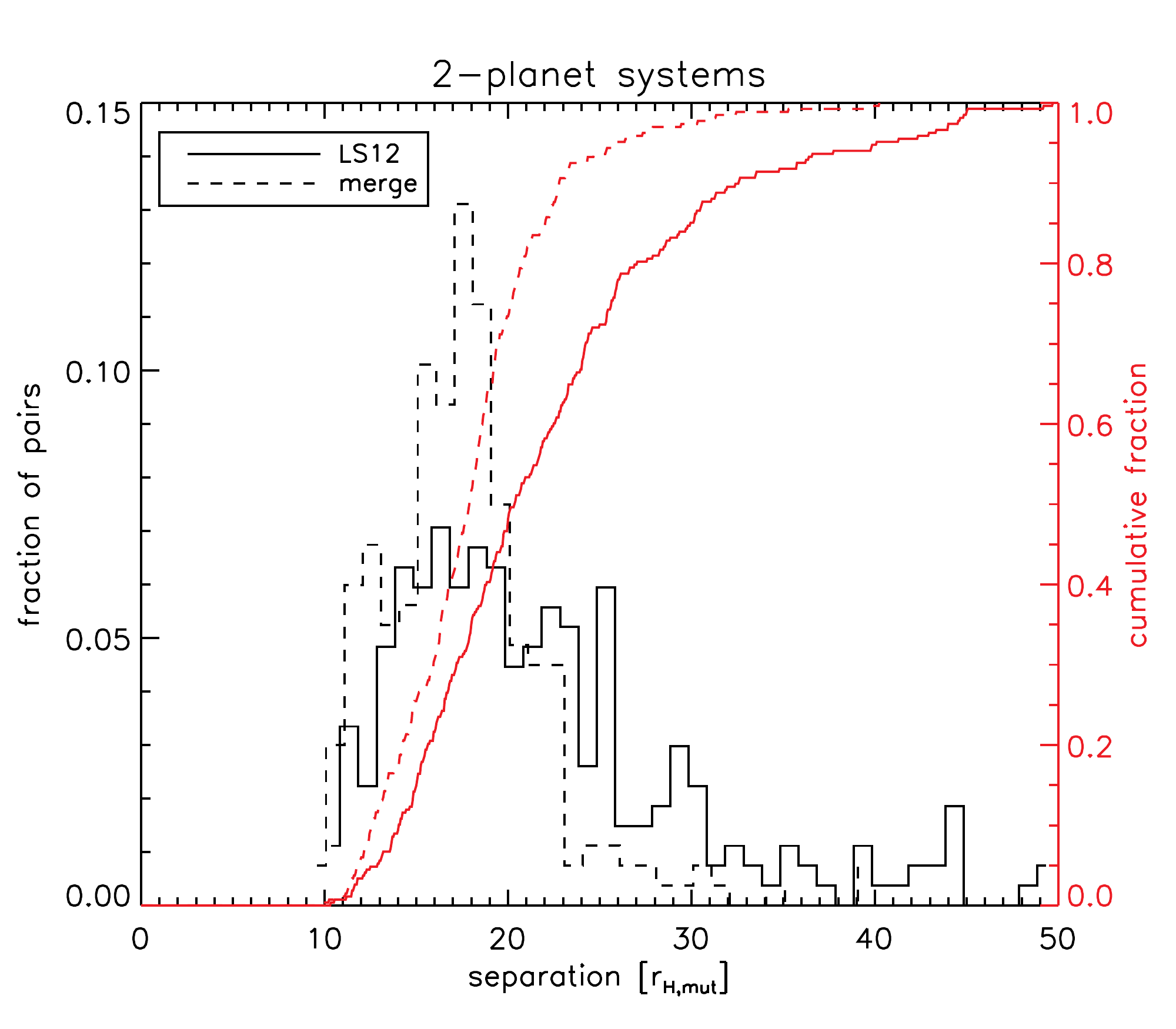}
  \includegraphics[width=0.5\textwidth]{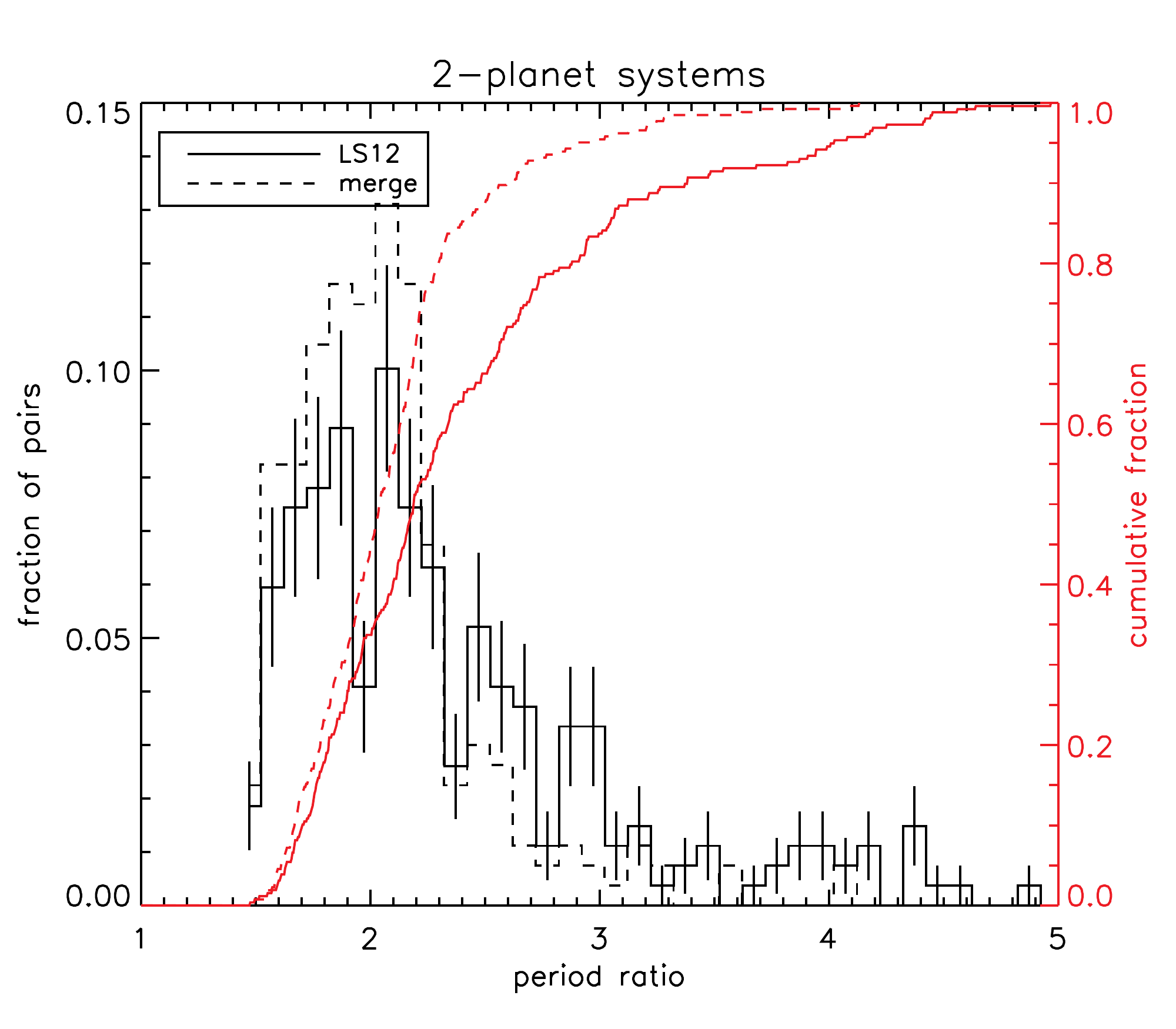}
  \caption{Separations of planets in 
    systems ending with two planets. Systems initially 
    had their inner planet at $0.1$\,au, with masses in the 
    range $1-10\mathrm{\,M}_\oplus$, and we include 
    all initial separations. Simulations run with the 
    LS12 collision algorithm are shown as a solid 
    line, and those with standard perfect merging as 
    a dashed line. The right-hand axes show the cumulative 
    distribution. \textbf{Top: }Separations in mutual 
    Hill radii. Assuming that all collisions result in 
    perfect merging yields a narrower and tighter range of 
    final orbital separations. \textbf{Bottom: }Period 
    ratios in the same systems. $1\sigma$ Poisson 
    error bars are marked on the solid histogram. 
    A marked $\sim4\sigma$ deficit of 
    planets near the 2:1 mean motion resonance is found 
    when using the improved collision algorithm.}
  \label{fig:delta-2p}
\end{figure}

Finally, in Figure~\ref{fig:delta-2p} we show the 
separations of surviving 
two-planet systems in the runs initially at 
$0.1$\,au. In the top panel we show the separations 
in mutual Hill radii. 
The runs with the LS12 collision 
algorithm result in much more widely-separated 
two-planet systems than those run with the 
perfect merging algorithm, and in fact the 
distribution more closely resembles the 
observed separations of \emph{Kepler} 
multi-planet systems which peaks at $\sim20$ 
mutual Hill radii \citep{Weiss+17}. In the lower 
panel we show the period ratios in the same systems. 
With the LS12 algorithm, we find a marked ($4\sigma$) deficit 
around a period ratio of 2, possibly associated to 
the 2:1 mean motion resonance (MMR); this feature persists 
independent of the bin size. This feature 
may arise because resonant and near-resonant orbits 
are less stable in the systems resulting from the LS12 runs 
than in those run with perfect merging: the planets in 
the former have a slightly higher mean eccentricity 
($0.13$ versus $0.08$) which could render the resonances 
less stable. This may contribute to the 
observed lack of \emph{Kepler} planets in MMRs 
\citep{Lissauer+11,Fabrycky+14}, 
although to establish this would require 
a treatment of tidal eccentricity damping 
\citep[e.g.,][]{DelisleLaskar14} and the damping from collisional 
debris \citep[similar to the planetesimal damping considered by][]{ChatterjeeFord15}, 
both of which could result in small changes to the planets' semimajor 
axes that might affect this feature.

\section{Numerical study II: instability induced by outer dynamics}

\label{sec:outer}

\begin{figure}
  \vspace{-0mm}
  \begin{center}
    \includegraphics[width=0.45\textwidth]{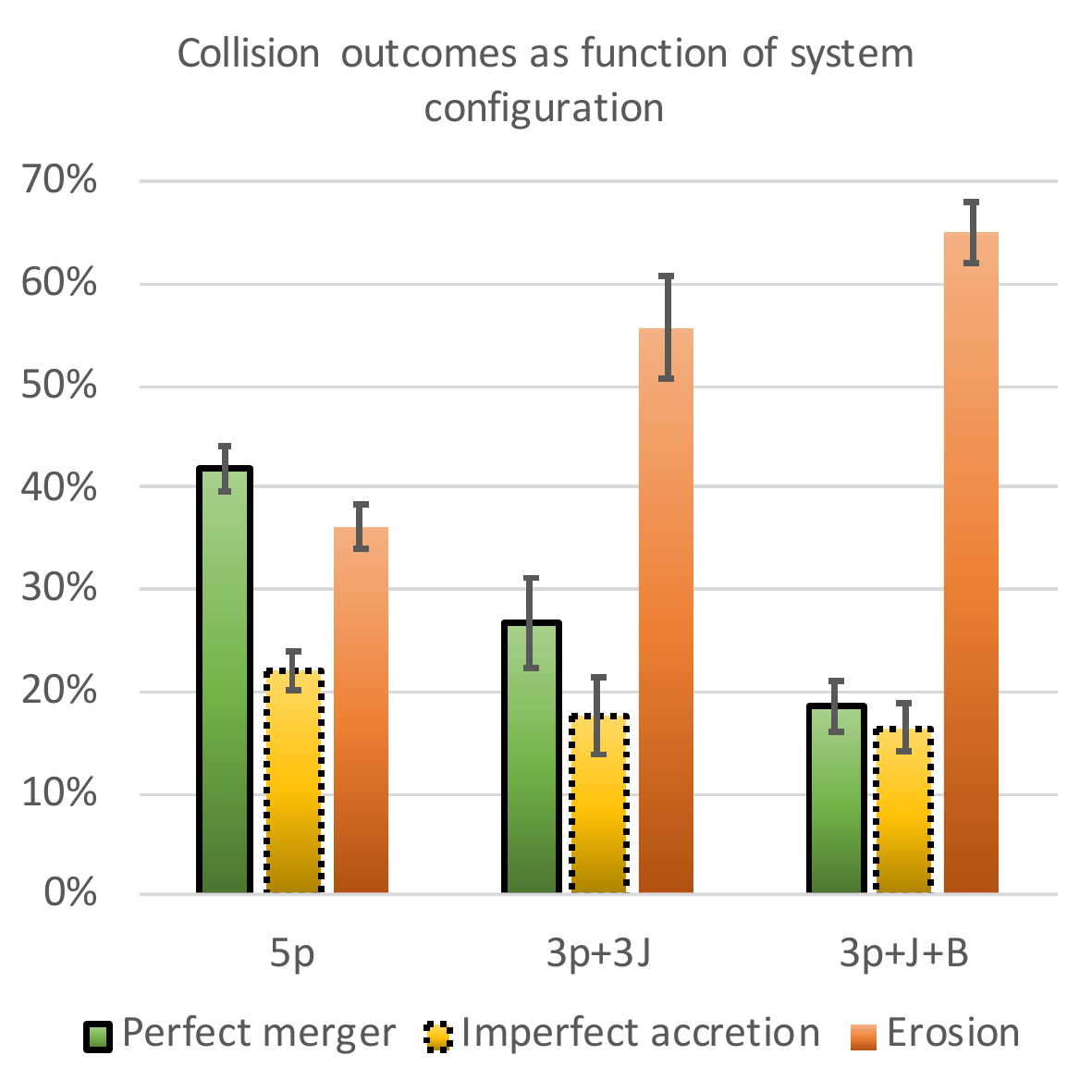}
    \includegraphics[width=0.5\textwidth]{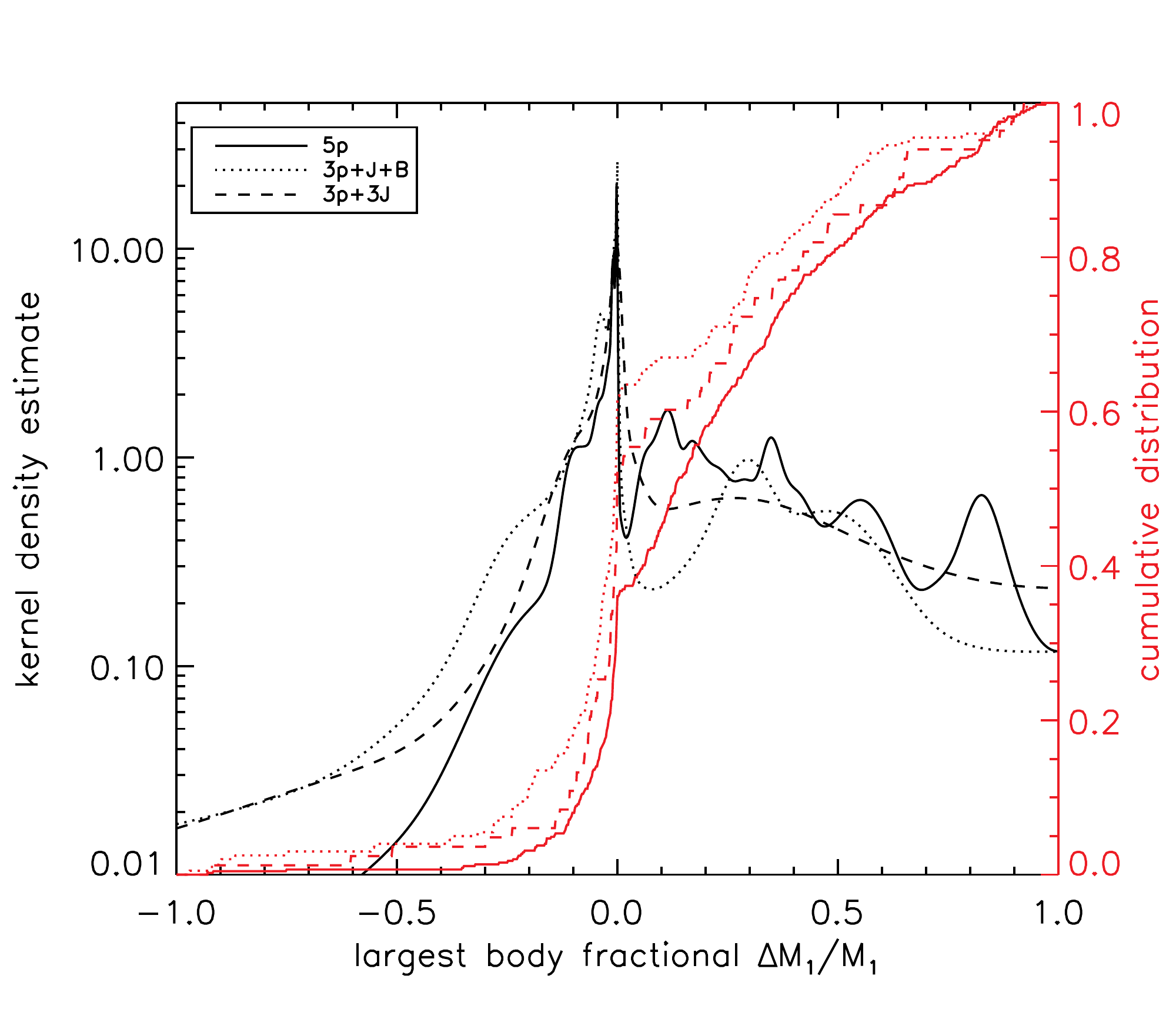}
  \end{center}
  \caption{Collision outcomes of self-unstable systems 
    (5p: the set of runs at $0.1$\,au from 
    Section~\ref{sec:inner}),  
    compared to those destabilised by dynamically 
    active outer systems: giant planets undergoing 
    Kozai perturbations (3p+J+B) and scattering (3p+3J). 
    \textbf{Top panel: }broad classification of 
    collision outcomes. \textbf{Bottom panel: }
    distribution of changes to largest body mass 
    in the collisions. 
    Planets in systems destabilised externally experience 
    even more violent collisions than those systems 
    that are intrinsically unstable.}
  \label{fig:dm-outer}
\end{figure}

In this section, we investigate the effects of changing the
collision model to the instabilities induced by an outer system, as
studied by \cite{Mustill+17}. As discussed in Section~\ref{sec:colls}, 
we expect that collisions will be more violent in these systems 
due to the large orbital eccentricities that external perturbers can 
excite.

\subsection{Setup}

In \cite{Mustill+17} we studied two dynamical scenarios: an inner system
together with an outer system of planets unstable to planet--planet
scattering (\textsc{Giants}), and an inner
system together with a single outer planet
and a wide binary stellar companion \textsc{Binaries}.
In either case, the excitation of eccentricity of the
outer planet(s) can lead to its pericentre approaching
or overlapping the orbits of the inner planets.
In roughly 1 in 4 of these
simulations, the inner system was destabilised by the outer system,
losing one or more of the inner planets. Planet--planet collisions
accounted for around half of the planets lost from the inner system
($43\%$ in \textsc{Binaries}, and $51\%$ in \textsc{Giants}).

Motivated by the results of \cite{Mustill+17}, we 
run simulations of three-planet systems with 
extra bodies in the outer system. For our inner architectures, 
we depart from the setup of \cite{Mustill+17}, 
who took actual triple-planet \emph{Kepler} systems 
as a template, as planets in these systems often have 
a significant gaseous component. Instead, we removed 
the second and fourth planets from the quintuple 
systems we constructed for Section~\ref{sec:inner}, 
leaving highly-spaced triple-planet systems.

We construct two sets of systems with different 
outer system architectures. For the first set 
(3p+3J), we add three Jupiter-mass planets 
with the innermost at 1\,au and the others separated 
by $4-6$ mutual Hill radii. Using equal-mass 
planets ensures very strong scattering that 
will be very disruptive for the inner system 
and efficiently excite eccentricities 
\citep{Carrera+16,Huang+17}. For the second set 
(3p+J+B) we add a single Jupiter-mass 
planet at 1\,au, and a Solar-mass binary at 40\,au 
\citep[the peak of the period distribution for 
  Solar-type stars,][]{DucheneKraus13} with an 
eccentricity of $0.2$ and an inclination of 
$50^\circ$, sufficient to drive Kozai cycles 
on the giant planet, but that restricts 
the planet's pericentre from sweeping through 
the entire inner planetary system and destroying 
all planets. We compare these systems to the 
fiducial 5-planet systems at $0.1$\,au from 
Section~\ref{sec:inner}, which we refer to 
as 5p in this section.

\subsection{Results}

The frequency of collision outcomes, and the distributions 
of the changes to the mass of the largest body, are 
shown in Figure~\ref{fig:dm-outer}. We also show on 
these plots the statistics from our fiducial self-unstable 
5-planet simulation set (at $0.1$\,au and with separations 
of $4-6r_\mathrm{H,mut}$). For these runs including outer 
planets, we only include in the statistics collisions where 
both of the planets were members of the inner triple system 
(or their collisionally accreted or eroded descendants), 
as the collision prescription we use is calibrated for 
rocky, not gaseous, planets.

The systems destabilised by the external planets experience 
far more erosive collisions than the self-unstable systems: 
only $20-30\%$ of the collisions in these runs resulted in 
perfect mergers, while $50-60\%$ were erosive. This compares 
to the $\sim35\%$ erosive collisions in the self-unstable 
systems. Individual erosive impacts are often more destructive: 
we find $2.4\%$ and $0.8\%$ supercatastrophic disruptions in 
3p+3J and 3p+J+B compared to only 
$0.4\%$ in the set 5p. In systems which experienced 
collisions between the inner planets, $13.8\%$ of the total 
planetary mass was lost in collisions in 3p+3B and 
$21.8\%$ in 3p+J+B, compared to only $11.4\%$ in 
5p. However, even in the extreme case of the binary-perturbed 
3p+J+B systems, the final distributions of planet masses 
in the destabilised systems are not statistically 
distinguishable ($p=0.29$ on a KS test, albeit with only 
around 30 planets in each sample).

\section{Numerical study III: \emph{In-situ} planet formation from embryos}

\label{sec:formation}

Motivated by the decreasing prevalence of perfect mergers 
at lower planet masses, 
we now study the effects of changing the collision prescription 
on \emph{in-situ} formation of super-Earths from rocky embryos. 
While several authors have now considered the effects of 
adopting a more realistic collision prescription in the 
formation of the terrestrial planets 
\citep{Chambers13,Carter+15,Leinhardt+15,Chambers16,Quintana+16}, 
studies of similar formation processes for close-in planets 
have so far mostly modelled collisions as perfect mergers 
\citep[][but see \citet{Wallace+17} for an exception]
{HansenMurray12,HansenMurray13,ChatterjeeTan14,Schlichting14,
MoriartyFischer15,Ogihara+15,MoriartyBallard16,Izidoro+17,
MatsumotoKokubo17}. 
This is despite the fact that the assumption that collisions always 
result in perfect merging becomes worse as one moves closer to the 
star, as we showed above. Here, we explore the effects of 
adopting the more realistic collision model.

We run simulations with 10 Earth masses of material in 100 
$0.1\mathrm{\,M}_\oplus$ embryos, with the innermost placed at 
$0.1$\,au and embryos spaced by $4-6$ mutual Hill radii. This 
sets the outermost 
embryo at around $1.7-1.9$\,au. Initial inclinations are up 
to $0.1^\circ$. We run these simulations for 10\,Myr with 
either the perfect merging algorithm, and with the improved 
algorithm based on LS12 with all collision fragments 
removed. We furthermore run sets of simulations starting 
from 100 smaller $0.05\mathrm{\,M}_\oplus$ embryos. 
There were 10 simulations in each set.

\begin{figure}
  \includegraphics[width=0.5\textwidth]{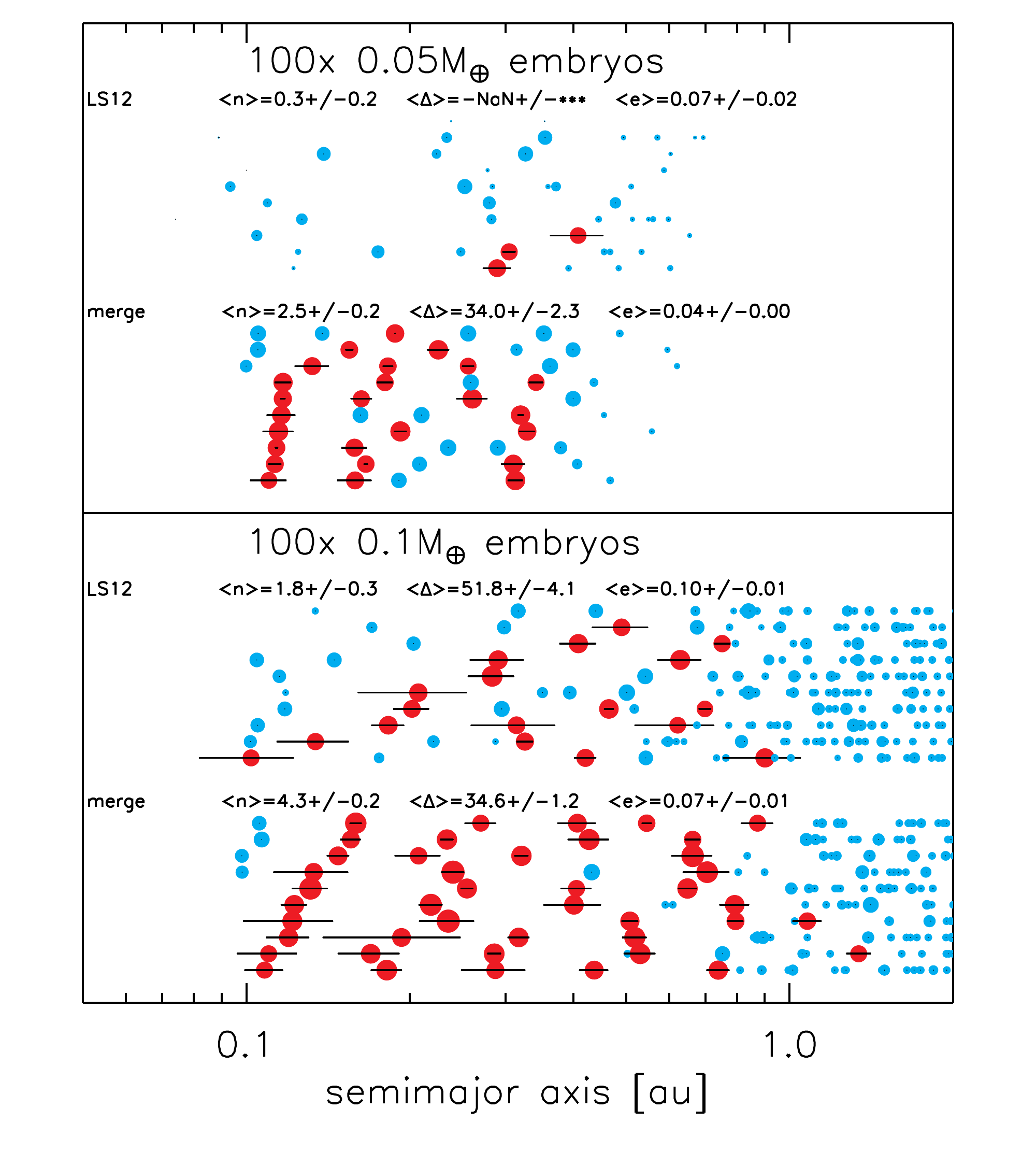}
  \caption{Effects of the collision algorithm on planet formation
    from Mars-sized embryos. In the upper panel we start with 100 
    $0.05\mathrm{\,M}_\oplus$ embryos; in the lower, 100 $0.1\mathrm{\,M}_\oplus$ 
    embryos. Systems are shown after 10\,Myr of evolution, by
    which point planet formation is still ongoing beyond $\sim1$\,au but
    has completed within a few tenths of an au.
    Symbol size is proportional to the cube root of mass, and
    planets more than at least $1\mathrm{\,M}_\oplus$ are in red, those smaller
    (whether embryos or fragments) in blue. Eccentricities are shown for
    the planets over $1\mathrm{\,M}_\oplus$ as horizontal lines. The lower
    set of systems in each panel are run with perfect merging, the upper set with the LS12
    collision prescription with fragments instantly removed. For each set, we show the mean number
    of planets at least Earth's mass, and the mean separations (in mutual Hill
    radii) and eccentricities of these planets (error bars show standard
    errors on the means). Perfect merging results in high-multiplicity systems;
    LS12 with fragments removed results in more widely-spaced, lower multiplicity
    systems, and when starting from the smaller embryos, planet formation 
    is almost entirely suppressed.}
  \label{fig:lovisplot}
\end{figure}

Simulation results are shown in Figure~\ref{fig:lovisplot}. 
Each row shows the semi-major axes of bodies in a single 
system. Red symbols show planets of at least one Earth mass, 
with horizontal bars showing the eccentricities of these large objects.
By the end of the simulation at 10\,Myr, planet formation
is complete out to $\sim1$\,au.

With the larger $0.1\mathrm{\,M}_\oplus$ embryos, 
when run with perfect merging, a large number of Earth-mass 
bodies form, out to around 1\,au (where planet formation is still ongoing). 
The final systems have a mean of $4.3$ planets per system, separated by a mean of 35 mutual 
Hill radii. This changes markedly if run with the LS12 algorithm, 
removing collision fragments: 
in these systems, an average of only 1.8 planets more massive than Earth 
form, and the multiple systems are more widely spaced (mean of 52 mutual 
Hill radii). We formed three intrinsic singletons and one intrinsic ``planetless'' star 
(its largest planet was $0.82\mathrm{\,M}_\oplus$). These simulations lost 
33\% of all their material, and 41\% of material within 1\,au. We 
also tested the effects of retaining the collision fragments in 
these systems. In these cases, fragments were quickly reaccreted, 
and the multiplicities of systems were again high (a mean of 
$3.4$ planets more massive than Earth).

The effects of the LS12 algorithm are more pronounced 
when starting from the smaller 
($0.05\mathrm{\,M}_\oplus$) embryos, owing to the higher impact 
velocities relative to escape velocities. When starting from 
one hundred $0.05\mathrm{\,M}_\oplus$ embryos with low inclinations, 
we formed one single-planet, three two-planet and six 
three-planet systems with perfect merging, but only three 
single-planet and seven zero-planet systems with LS12, removing 
the fragments. Adopting the more realistic collision algorithm 
can thus strongly curtail \emph{in-situ} planet formation 
from small embyos, so long as the collisional debris is 
quickly removed from the system.

Our simulations transition rapidly 
  from a majority forming no planets (when starting from 
  $0.05\mathrm{\,M}_\oplus$ embryos) to a majority forming 
  multiple planets (starting form $0.1\mathrm{\,M}_\oplus$ embryos). 
  Although we suffer from small-number statistics, this may 
  suggest a critical embryo mass above which embryos at $\sim0.1$\,au 
  can avoid collisional grinding and consistently grow to larger planets. It may 
  also raise a fine-tuning problem with this model, since one 
  interpretation of the statistics of transit
  multiplicities from the \emph{Kepler} mission is that
  half or more of planetary systems contain only one
  planet within $\sim1$\,au \citep[e.g.,][]{Johansen+12,BallardJohnson16}, 
  although this interpretation is not unique \citep[see e.g.,][who favour
  a larger fraction of multiple systems whose observed transit multiplicities 
  are reduced by inclination excitation]{Zhu+18}.

We note that a recent study by \cite{Wallace+17} 
found that fragmentation does \emph{not} impede the formation 
of rocky planets on short orbital periods. \cite{Wallace+17} 
used a collision prescription similar to ours, also based on 
the results of \cite{LeinhardtStewart12}, with some minor 
differences in the treatment of hit-and-run collisions. The 
major difference, however, is in the treatment of smaller 
collision fragments: while we remove these, \cite{Wallace+17} 
retained them, with the qualification that objects could not 
be reduced below a ``Minimum Fragment Mass''. As we discuss in 
Section~\ref{sec:fragments} below, we expect most fragments 
to be ground to smaller sizes in a collisional cascade, 
and then removed by radiation forces,  
before they re-accrete onto one of the larger planets. 
By imposing a Minimum Fragment Mass, \cite{Wallace+17} 
did not capture this collisional cascade. Our study and that of 
\cite{Wallace+17} thus represent the two extreme cases of 
no mass removal and instant mass removal. Further study 
would require collisional modelling of the debris fragments 
within the $N$-body simulations, but we suspect that reality 
would lean more towards our implementation at smaller orbital 
distances ($\sim0.1$\,au) and more towards that of \cite{Wallace+17} at larger 
orbital distances ($\sim1$\,au), where orbital velocities 
are lower, collisional velocities are lower, 
 and gravitational focusing enhances re-accretion.

\section{Discussion and conclusions}

\label{sec:discussion}

\subsection{Implications for the \emph{Kepler} dichotomy}

The ``\emph{Kepler} dichotomy'' refers to the large number 
of systems possessing only one transiting planet 
compared to multiple systems 
(\citealt{Johansen+12}; see also 
\citealt{Lissauer+11,FangMargot12,TremaineDong12}). 
Two possible causes of the excess of singles are 
formation (some systems just form with only one 
planet within a few tenths of an au) and later 
dynamical evolution (many or all systems form as 
multiples, and most are unstable and reduce to 
single-planet systems). 

Attempts to reproduce 
the dichotomy through dynamical instabilities 
have met with mixed success. \cite{Mustill+17} 
and \cite{Huang+17} show the strong effects 
instabilities amongst outer planets can have 
on inner planets, but the relatively small occurrence 
rate of gas giants means that this can only make 
a modest contribution to the destabilisation of 
the population of transiting systems (in 
\citealt{Mustill+17}, we argued that $\lesssim20\%$ 
of multiple systems will be destabilised by 
outer planets, with the uncertainty dominated by 
the poorly-constrained occurrence rate of sub-Jovian 
planets beyond 1\,au). The recent work by \cite{Izidoro+17} 
finds that resonant chains of planets are frequently 
unstable, although still not frequently enough to match the 
\emph{Kepler} data. 

Our simulations show that the incorporation of 
an improved collision model does not have a significant 
effect on the multiplicities of systems after instability. 
Based on the weak sensitivity of collision outcomes to the 
initial spacings in our non-resonant 5-planet systems, 
we expect that the effects of changing the collision algorithm 
in unstable resonant systems will not be significantly 
different. We also do not form intrinsically zero-planet 
systems through continued grinding down of the planets, as 
hypothesised by \cite{VolkGladman15}. Even when perfect 
merging is abandoned, most collisions between mature planets 
more massive than Earth do not result in enough 
mass loss to significantly affect the planetary mass 
distribution or the number of large, detectable planets.

Observed multiplicity is a function not only of a system's 
intrinsic multiplicity but of the planets' separations 
and mutual inclinations. Our systems evolved with the LS12 
algorithm end up slightly more widely spaced and dynamically 
excited than those run with perfect merging. This has a moderate 
effect on observed multiplicities: taking the systems in our 
baseline simulations ($a=0.1$\,au), we clone the systems at the 
end of the simulation to generate a large sample of 10\,000 
systems and observe these from random orientations. The 
ratio of systems observed with two transiting planets to one 
transiting planet is $2.5:1$ for the simulations with perfect merging 
and $3.4:1$ for those run with the LS12 algorithm, despite the 
intrinsic multiplicities being almost identical. This falls far 
short, however, of the observed ratio of around 6:1 
\citep[e.g.,][]{Johansen+12,Lissauer+14}.

The alternative explanation for the \emph{Kepler} 
dichotomy---that it arises from the processes of 
planet formation---fares slightly better from our results. 
Previous attempts to explain the dichotomy through 
\emph{in-situ} accretion of embryos have invoked 
a large range of surface density gradients in the planetesimal disc 
\citep{MoriartyBallard16}. In our simulations, 
when abandoning the assumption of perfect merging 
between embryos, a greater diversity of outcomes 
is seen than when retaining perfect merging, with 
many simulations forming only one or even zero large 
($>1\mathrm{\,M}_\oplus$) planets, particularly when 
the initial embryo mass is smaller ($0.05\mathrm{\,M}_\oplus$). 
This is true 
so long as collision fragments are swiftly removed from 
the system and do not reaccrete. As pointed out 
  above, however, there is a potential fine-tuning issue 
  in getting enough single-planet systems in these simulations.

\subsection{The fate of the collision debris}

\label{sec:fragments}

\begin{figure}
  \includegraphics[width=0.5\textwidth]{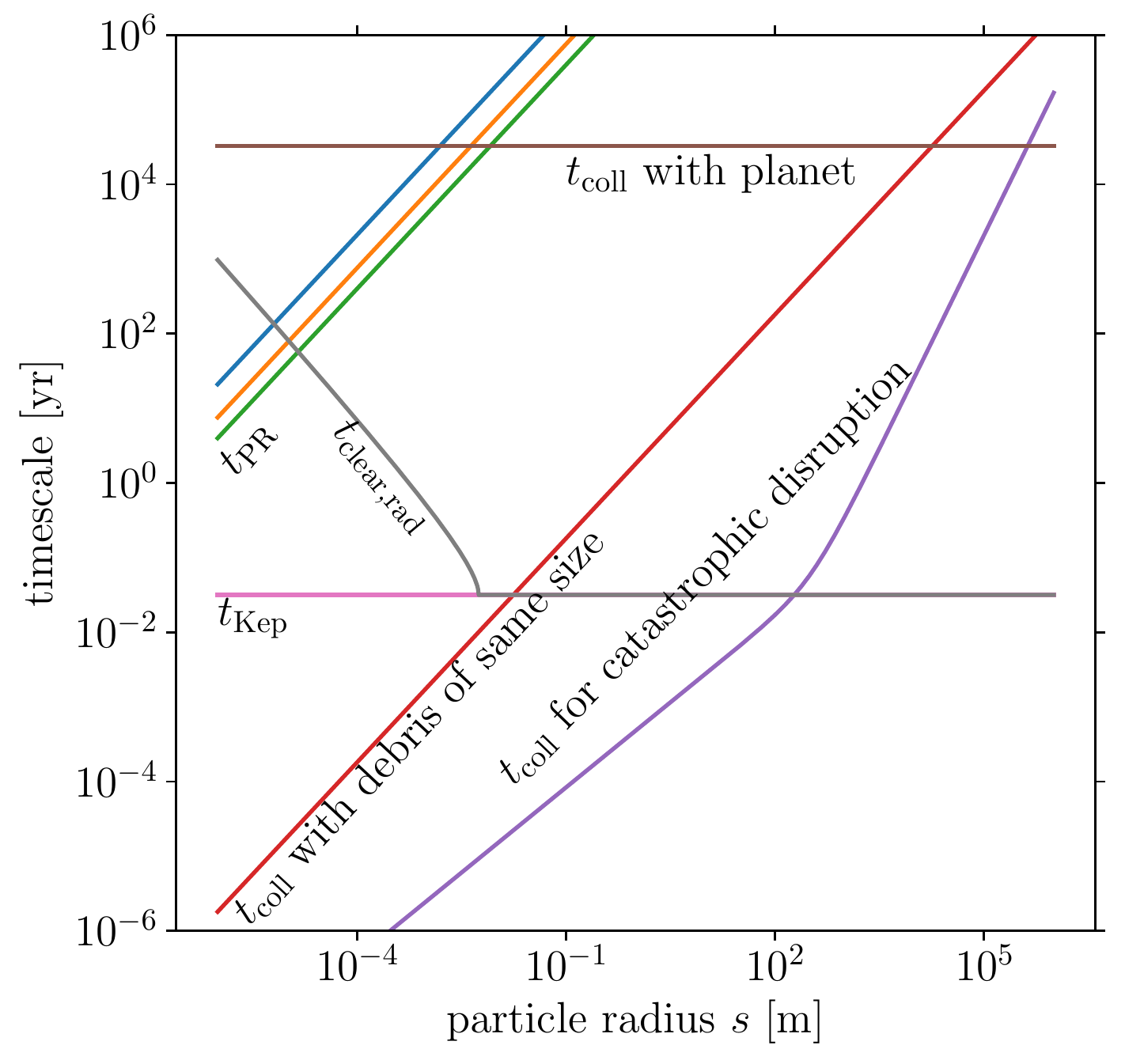}
  \caption{The fate of collision debris 
    produced in a collision at $0.1$\,au. At this distance, the 
    timescale for re-accretion onto a $1\mathrm{\,M}_\oplus$ 
    planet is almost $10^4$\,yr (brown line). This is shorter than the 
    lifetime under PR drag for particles $\gtrsim1$\,mm in 
    size (diagonal green, orange, blue lines show the time for 
    particle orbits to shrink 10, 20 and 100\%). 
    However, these particles will collide and be destroyed by 
    other debris particles on still shorter timescales, less 
    than one orbital timescale for bodies $\lesssim1$\,cm in size 
    (Diagonal red line shows the time to collide with another 
    debris particle if the size distribution is monodisperse; 
    bent purple line shows the time to experience a catastrophic 
    disruption with a smaller debris particle).
    A collisional cascade would then quickly reduce these particles 
    to the blow-out limit ($\sim1\,\mu\mathrm{m}$), where they 
    are removed from the system by radiation pressure on a dynamical 
    time-scale (horizontal pink line). 
    If a significant number of small particles are generated, 
    however, the disc becomes optically thick. Blow-out grains 
    are only removed if they exist on high-inclination orbits 
    where the optical depth to the star $\tau_\mathrm{rad}\lesssim1$. 
    The time-scale for this process, 
    $t_\mathrm{clear,rad}$ (grey line), is 
    still shorter than the time-scales for re-accretion by 
    the planet and for PR drag to remove the particles.}
  \label{fig:timescales}
\end{figure}

We now revisit the fate of the collision debris. 
  Debris is removed from systems by several processes. 
  Small grains ($\lesssim1\mathrm{\,\mu m}$ for a G-type star) are 
  removed by radiation pressure on a dynamical time-scale, 
  while larger grains experience orbital decay through Poynting--Robertson 
  drag \citep{Burns+79}. They are also reduced to smaller sizes through destructive 
  collisions; the subsequent collisional cascade results in 
  the particles eventually being ground small enough to be removed by 
  radiation pressure. The key question is whether these removal 
  methods occur faster than the parent planets will re-accrete 
  the debris.

The debris will emerge from the collision with a 
  certain size distribution. In their numerical experiments, 
  \cite{LeinhardtStewart12} found a very steep size distribution with most 
  of the mass in the smallest fragments; in terms of diameter $D$,
\begin{equation}
  n(D)\textrm{d}D = CD^{-(\beta+1)}\mathrm{d}D
\end{equation}
where $\beta$ ranges from $2.5$ to $5.2$ with 
  a median of $3.8$ (their Equation~31 and Table~1). 
  Furthermore, as well as larger gravitationally-bound 
  fragments, some ejecta will be in the form of 
  small melt droplets: \cite{Benz+07} found
  that the size distribution of ejecta peaked at $s\sim1$\,cm
  in their simulations of collisional stripping of Mercury's mantle. 
  Here we initially assume a monodisperse population 
  of debris fragments and then consider a more realistic 
  continuous distribution. In general, the more the size distribution 
  is weighted towards smaller particles, the shorter 
  the lifetime of the debris.

Consider a collision at $a=0.1$\,au that creates a monodisperse population 
  of grains with radius $s$ and the combined mass of a body of 
  radius $s_\mathrm{eq}=1000$\,km, and that leaves a $1\mathrm{\,M}_\oplus$, 
  $1\mathrm{\,R}_\oplus$ planet at $0.1$\,au. Assume that the grains have a 
  mean orbital eccentricity of $\langle e\rangle =0.2$. The time-scale for reaccretion 
  onto the planet, including the effects of gravitational focusing (fairly unimportant 
  at these orbital velocities, increasing the gravitational cross-section by only $35\%$), 
  is then $\sim33\,000$\,yr. This is shown as the horizontal brown 
  line in Figure~\ref{fig:timescales}. For our neglect of debris to be justified, the particles must be 
  removed on shorter timescales.

Stellar radiation causes a slow decay of particle orbits through 
the Poynting--Robertson effect (``PR drag''), which causes a low-eccentricity 
orbit to decay into the star in a time
\begin{equation}
 t_\mathrm{PR} = a^2/4\alpha
\end{equation}
where
\begin{equation}
  \alpha = \frac{3L_\star}{16\pi c^2 \rho s},
\end{equation}
$L_\star$ being the stellar luminosity and $\rho$ the density 
of a debris particle. 
For our particles, it suffices to drift a distance of 
$\sim a\langle e \rangle$ to prevent reaccretion, leading to 
a slightly shorter lifetime (diagonal green, orange and blue lines in 
Figure~\ref{fig:timescales}). 
This is shorter than the reaccretion timescale for 
mm- to cm-sized particles, and these particles will be removed 
without significant re-accretion onto the planet. Hence, ignoring 
collisions between debris fragments, 
recondensed vapour droplets are lost to PR drag before re-accretion.

However, owing to the large density of particles, the 
  lifetimes of particles of all sizes are in fact 
  also limited by mutual collisions. 
  Considering a monodisperse size distribution, in 
  which every collision is with an equal-mass object and is 
  destructive, the particles have a collision lifetime of 
\begin{equation}
  t_\mathrm{coll} = \frac{t_\mathrm{Kep}}{\pi\tau_\mathrm{eff}},
\end{equation}
where $\tau_\mathrm{eff}$ is the effective face-on optical depth 
of the annulus of fragments. $t_\mathrm{coll}$ is 
shown as the diagonal red line in Figure~\ref{fig:timescales}. 
This is in fact less than the orbital timescale for 
grains $\lesssim1$\,cm in size, and is less than 
the timescale for re-accretion onto the planet for all but 
the largest fragments of $s\gtrsim10$\,km: the surface area of the fragments 
is always large, and the boost to the planet's 
gravitational cross-section through gravitational focusing 
is modest because of the high orbital velocities. 
Thus, particles will grind down to the blow-out size and be 
removed by radiation pressure, just as in typical 
debris discs \citep[e.g.,][]{Wyatt05}.

Considering a more realistic size distribution 
can shorten the collision lifetimes among the debris 
particles significantly. The reason is that a catastrophic 
disruption of a large fragment can be induced by a 
significantly smaller impactor. Assuming for simplicity a bimodal 
size distribution with equal mass in particles of size $s$ 
and the smallest particles sufficient to disrupt them 
yields a lifetime for the larger particles given by the 
purple line in Figure~\ref{fig:timescales}. The kink at $\sim100$\,m is 
caused by the change from the gravity-dominated to the 
strength-dominated regime. These timescales are now 
smaller than the timescale for re-accretion 
onto the planet, or comparable to in the case of the largest 
debris fragments (a few hundred km). However, recall that \cite{LeinhardtStewart12} 
found that in most cases the fragment distribution was 
bottom-heavy. Hence, most debris will be ground down to the blow-out 
size before re-accretion onto the planet.

Finally, the large mass of the debris cloud does however introduce 
a complication: if most of the grains are $\lesssim1\mathrm{\,cm}$ 
in size, the disc becomes optically thick and radiation pressure 
will be inefficient as an agent of removal. In this case, 
only grains on high-inclination orbits that can escape the 
disc mid-plane to less dense high-$z$ regions, where the radial optical 
depth $\tau_\mathrm{rad}\lesssim1$, 
will be removed. If we assign the grains a Rayleigh distribution 
of inclinations with parameter $\sigma_I$ (in radians), 
then the optically thin surface is at 
\begin{equation}
z_\mathrm{thin} = \sqrt{-2a^2\sigma_I 
  \log\left[2^{3/2}\pi^{1/2}\sigma_I(a/s)^2/N_\mathrm{part}\right]},
\end{equation}
where $N_\mathrm{part}$ is the number of particles.
There is then a fraction
\begin{equation}
f_\mathrm{thin} = 2\mathrm{sf}\left(z_\mathrm{thin}/a\sigma_I\right)
\end{equation}
of grains in the region with $\tau_\mathrm{rad}\lesssim1$, 
where $\mathrm{sf}(x)$ is the Gaussian survival function. 
These are removed on a dynamical time-scale, after which 
the next surface layer of grains becomes exposed; the erosion 
therefore takes place on a timescale
\begin{equation}
t_\mathrm{clear} = t_\mathrm{Kep}/f_\mathrm{thin}.
\end{equation}
For our parameters, the removal timescale is equal to the orbital 
timescale for the optically thin discs with $s\gtrsim1$\,cm, 
and rises to $\sim1000$\,yr if all of the mass is in small grains. 
This is still considerably lower than the timescale for re-accretion 
onto the planet, justifying our neglect of the debris. This timescale 
is plotted in Figure~\ref{fig:timescales} as the grey line.

We repeated these calculations for an orbital radius of $1$\,au. 
  Here the size where the mutual collision timescale exceeds that 
  for re-accretion onto the planet falls to 100\,km. We also repeated 
  the calculations at $0.1$\,au but with half the $e$ and $i$ of the fragments. 
  This reduced the timescale to collide with the planet, and slightly increased 
  the lifetime of debris particles to mutual collisions, but the mutual collsion 
  lifetime remained shorter than the re-accretion time for bodies under a few 
  hundred km.

\subsection{Conclusions}

We have implemented a collision algorithm into 
\textsc{Mercury} that improves on the 
standard algorithm that collisions between 
planets always result in perfect merging. We 
tested the effects of this on the outcome of 
$N$-body integrations of unstable multi-planet 
systems close to the star, systems destabilised 
by outer planets, and \emph{in-situ} formation 
of rocky super-Earths. In general, 
the effects of adopting the improved collision algorithm are 
greater when the ratio of the collision velocity to the 
escape velocity rises, such as when planetary orbits 
are smaller, eccentricity ecitation is stronger, or planets 
are of lower mass. Specifically, we find the following:
\begin{itemize}
\item Collisions between transiting planets at $\sim0.1$\,au 
  are frequently erosive. Perfect mergers only account for 
  $40\%$ of collisions in our fiducial case of an unstable 
  system of five super-Earths where the innermost is located 
  at $0.1$\,au (Figure~\ref{fig:dm}).
\item The fraction of mass lost to collisional debris 
  in these systems is $\sim10\%$ of the initial 
  planetary mass. Hence, the mass distribution and system multiplicity after 
  instability is not strongly affected compared to running 
  simulations with only perfect merging (Figure~\ref{fig:npl}). However, 
  the collision algorithm does have some affect on 
  the final separations of planets: surviving two-planet 
  systems are more widely spaced with the improved collision 
  algorithm than when run with the standard algorithm 
  (Figure~\ref{fig:delta-2p}).
  Furthermore, we can expect the distribution of 
  planetary radii to be more strongly affected 
  due to the strong dependence of planet radius on 
  envelope mass for planets with small hydrogen/helium 
  envelopes.
\item The fraction of collisions resulting in perfect mergers 
  rises as planets' semimajor axis is increased and decreases 
  as planets' mass is decreased. It is fairly insensitive to the 
  initial separation of planets (in mutual Hill radii). There is 
  a small effect when starting from 
  very flat configurations $i\sim0.1^\circ$ (Figures~\ref{fig:dm}, 
  \ref{fig:outcomes-i0.1}).
\item Transiting systems destabilised by eccentric outer bodies 
  (giant planets experiencing Kozai cycles or planet--planet 
  scattering) experience a smaller fraction of perfect mergers 
  ($20-30\%$), and the effects of a realistic collision prescription 
  are more significant in such systems (Figure~\ref{fig:dm-outer}).
\item Smaller planets, or planetary embryos, suffer more from 
  erosive collisions. This affects the outcome of planet 
  formation \emph{in situ} from smaller embryos. If collisional 
  debris is rapidly removed from the system (for example, by 
  radiation forces) then \emph{in-situ} formation forms fewer, 
  more widely-spaced planets with the improved collision algorithm 
  than when assuming that all collisions result in perfect merging 
  (Figure~\ref{fig:lovisplot}. This may provide a contribution 
  towards explaining the \emph{Kepler} dichotomy, at least for 
  smaller, rocky, planets. However, if debris is not quickly removed,  
  then reaccretion will lead to higher-multiplicity systems 
  similar to those formed if collisions always result in 
  perfect mergers.
\end{itemize}

\section*{Acknowledgements}

The authors are supported by the project grant 2014.0017 ``IMPACT'' 
from the Knut and Alice Wallenberg Foundation. 
A.J.\ was supported by the European Research Council under ERC 
Consolidator Grant agreement 724687-PLANETESYS, the Swedish Research
Council (grant 2014-5775) and the Knut and Alice Wallenberg Foundation 
(grants 2012.0150, 2014.0017, and 2014.0048).
The simulations were performed on resources provided by the Swedish 
National Infrastructure for Computing (SNIC) at Lunarc.
This research has made use of the Exoplanet Orbit Database
and the Exoplanet Data Explorer at \url{exoplanets.org}.
We thank the referee, John Chambers, for a careful 
and insightful referee report.

\bibliographystyle{mnras}
\bibliography{3p+p-inner}

\bsp    % typesetting comment
\label{lastpage}

\end{document}